\documentclass[12pt,preprint]{aastex}

\def\geqsim{\lower.73ex\hbox{$\sim$}\llap{\raise.4ex\hbox{$>$}}$\,$}
\def\leqsim{\lower.73ex\hbox{$\sim$}\llap{\raise.4ex\hbox{$<$}}$\,$}
\def\Mpch {~h^{-1}~{\rm Mpc}}

\newcommand{\myemail}{aross2@uiuc.edu}

\newcommand{\SDSSpixt}{{\em SDSSpix~}}
\newcommand{\SDSSpixb}{{\em SDSSpix}}

\slugcomment{{\apj~accepted May 30th, 2006}}
\shorttitle{Precise Higher-Order Angular Galaxy Correlations}
\shortauthors{Ross et al.}

\begin{document}

\title{Precision Measurements of Higher-Order Angular Galaxy Correlations Using 11 Million SDSS Galaxies}

\author{Ashley J. Ross\altaffilmark{1}, Robert J. Brunner\altaffilmark{1,2},  Adam D. Myers\altaffilmark{1,2}}

\email{\myemail}

\altaffiltext{1}{Department of Astronomy,Ê 
University of Illinois at Urbana-Champaign,Ê 
Urbana, IL 61801}
\altaffiltext{2}{National Center for Supercomputing Applications,
Champaign, IL 61820}

\begin{abstract}
 We present estimates of the N-point galaxy area-averaged angular correlation functions $\bar{\omega}_{N}$($\theta$)  for $N$ = 2,...,7 from the third data release of the Sloan Digital Sky Survey (SDSS).  The sample was selected from galaxies with $18 \leq r < 21$, and is the largest ever used to study higher-order correlations.  The measured $\bar{\omega}_{N}$($\theta$) are used to calculate the projected, $s_{N}$, and real space, $S_{N}$, hierarchical amplitudes.  This produces highly-precise measurements over 0.2 to 10 $h^{-1}$ Mpc, which are consistent with Gaussian primordial density fluctuations.  The measurements suggest that higher-order galaxy bias is non-negligible, as defining $b_{1} = 1$ yields $c_{2} = -0.24 \pm 0.08$.  We report the first SDSS measurement of marginally significant third-order bias, $c_{3} = 0.98 \pm 0.89$, which suggests that bias terms may be significant to even higher order.  Previous measurements of $c_{2}$ have yielded inconsistent results.  Inconsistencies would be expected if different data sets sample different galaxy types, especially if different galaxy types exhibit different higher-order bias.  We find early-type galaxies exhibit significantly different behavior than late-types at both small and large scales.  At large scales ($r > 1~h^{-1}$ Mpc), we find the $S_{N}$ for late-type galaxies are lower than for early-types, implying a significant difference between their higher-order bias.  We find $b_{1,early} = 1.36 \pm 0.04$, $c_{2,early} = 0.30 \pm 0.10$, $b_{1,late} = 0.81 \pm 0.03$, and $c_{2,late} = -0.70 \pm 0.08$.  Our results are robust against the systematic effects of reddening and seeing.  The latter introduces minor structure in $\bar{\omega}_{N}$($\theta$). 

\end{abstract}
\section{Introduction}
 A significant portion of our cosmological knowledge comes from studying the statistics of the density field traced by observable structure --- most often the temperature anisotropies of the cosmic microwave background (CMB) or the distribution of galaxies.  Many of these studies use the two-point correlation function (2PCF).  This is largely due to the fact that this statistic is relatively easy to calculate and model.  The 2PCF completely characterizes a Gaussian density field, as odd higher-order moments vanish and even moments can be entirely expressed in terms of the two-point function.  Using first-order linear perturbation theory (LPT) one can show that a Gaussian primordial distribution of density fluctuations will evolve under gravity into a distinctly asymmetric density field.   The observed higher-order moments of the local density field are predicted to  display hierarchical scaling with constant hierarchical amplitudes, $S_{N}$ (see equation 4, or \citealt{P80}).  

In the dynamical range where structure grows linearly or quasi-linearly (i.e. larger than the typical size of clusters) the $S_{N}$ are independent of cosmic epoch.  Thus, when using galaxies to measure $S_{N}$, any departure from the stated hierarchy can be due only to a non-Gaussian primordial density field or scale dependent bias between galaxies and the underlying density field.  Further, LPT predicts specific values for the $S_{N}$ and relates these values to the measured $S_{N}$ for galaxies and the bias.  Thus, measuring the higher-order correlation functions of the local density field to high precision can constrain the conditions of the primordial density field and the bias between galaxies and the underlying matter distribution.

Early measurements of $S_{N}$ suggested that the amplitudes are non-zero and support the hierarchical model on small scales (e.g., \citealt{Gro77, Saund91, Bouch93, Gaz94}).  More recent measurements endeavor to constrain $S_{N}$ on larger scales:  Using just over 200,000 galaxies from the 2dF Galaxy Redshift Survey (2dFGRS) at a median redshift of $z \sim$ 0.11, \cite{Cr04} found (after removing two major super-clusters from their field)  approximately constant $S_{N}$ for $N\leq5$ to 10 $h^{-1}$ Mpc.  Most recently, using just under 651,000 galaxies selected from the 2 Micron All Sky Survey (2MASS) with $K_{s} < $13.5 at a median redshift of $z =$ 0.074, \cite{FR05} found constant $S_{N}$ for $N \leq$ 7 to 30 $h^{-1}$ Mpc.  This result strongly supports Gaussian initial conditions.  Their measurements, however, also allowed them to determine the second-order bias parameter, $c_{2}$, to be 0.57$\pm$0.33 for $K_{s}$ band galaxies, only barely consistent with zero higher-order modes for galaxy bias and of opposite sign to previous results.  This measurement was based on their previous measurement of $b_{1} = 1.39 \pm 0.12$ using the 2MASS  galaxy angular power-spectrum \citep{FR05b}.

Separate recent measurements of two-point and three-point correlations using optical surveys have constrained the first- and second-order bias parameters.  By calculating the redshift space three-point correlation function, \cite{Gaz05} found $b_{1} = 0.94_{-0.11}^{+0.13}$ and $c_{2} = -0.36_{-0.09}^{+0.13}$ for $b_{j} < 19.45$ galaxies in the 2dFGRS ($c_{2} = \frac{b_{2}}{b_{1}}$).  \cite{Pan05} found, by studying the monopole contribution to the 2dFGRS three-point correlation function, that, for galaxies of absolute magnitudes $b_{j}$ between -21 and -20, $b_{1} = 1.04_{-0.09}^{+0.23}$ and $b_{2} = -0.06_{-0.001}^{+0.003}$.  \cite{H05} found, using the Sloan Digital Sky Survey (SDSS) spectroscopic galaxy catalog to calculate the bispectrum, that $c_{2}$ is consistent with zero to within 10$\%$.  Comparing early measurements of optical and IRAS galaxies, \cite{Fry93} determined that there exists both significant second-order bias and third-order bias between the two samples, suggesting that the measured second- and third-order bias terms may be dependent on galaxy type (as an infrared survey will certainly select galaxies of different spectral properties than an optical survey).

The measurement of $c_{2}$ by \cite{FR05} is somewhat puzzling in light of measurements using optical surveys.  The authors offer the explanation that the 2MASS data set likely contains a much larger fraction of early-type galaxies than typical optical surveys.  It is thus becoming important to determine the relationship between bias and galaxy type.  It has been shown (e.g., \citealt{W98,Z02,N02,Ma03}) that early-type galaxies cluster more strongly than other galaxy types, leading to a larger $b_{1}$ for early-type galaxies.  By calculating $S_{3}$ for both early- and late-type galaxies and comparing the results to those for the full sample of galaxies, one can determine the dependence of  $b_{1}$ and $c_{2}$ on galaxy type.  Of course, subdividing a sample by galaxy type becomes statistically more meaningful in larger data sets, such as that provided by the SDSS third data release (DR3).

Previously, \cite{Sza02} measured third- and fourth-order correlations for galaxies in the SDSS Early Data Release (EDR) and found that these measurements should be highly free of any systematics.  In this paper, we calculate and analyze the area-averaged angular correlation functions using SDSS DR3 galaxies, up to seventh-order.  Angular surveys have several advantages to redshift surveys when calculating high-order correlations: (1) they include vastly more objects;  (2) redshift distortions are not a problem (see, e.g., \citep{K87, H92}); and (3) they are not susceptible to rare peaks in the galaxy density field (see, e.g., \citealt{Cr04,Nic06}).  These factors, coupled with the area and depth of the SDSS DR3 should allow precise measurements of $S_{N}$ and their dependence on galaxy type.  This in turn will allow us to determine $c_{2}$ and its dependence on galaxy type.  After discussing our data in \S2 and techniques in \S 3 and \S 4, we present our measurements of the higher-order galaxy correlations for  early-type, late-type, and all galaxies in \S 5.  In \S 6 we discuss implications for linear and non-linear biasing schema and its dependence on galaxy type.  Throughout our analysis, we assume a flat $\Lambda$CDM cosmology (e.g., \citealt{Sper03}).
%\includegraphics[]{../SEDSzphot.jpg}

% NOTE FOR ROBERT: standard seems to be to drop the \primes on the photometric bands these days
\section{Data}
The data for these measurements were taken from the SDSS DR3 \citep{Ab}.  This survey obtains wide-field CCD photometry \citep{C} in five passbands ($u,g,r,i,z$; e.g., \citealt{F}).  The completed SDSS should cover $\sim$10,000 square degrees.  DR3 alone covers just over half of that area (5282 square degrees).  We selected galaxies with positions lying in the northern, contiguous portion of the SDSS from a DR3 {\tt PhotoPrimary} database and further constrained the sample (using the \citealt{Sc} dust maps) to have reddening-corrected magnitudes in the range $18 \leq r <  21$.  Further, significant masking was required to account for bright stars, areas of high reddening and poor seeing, (as discussed in detail in section 4.)  This produces a set of just over 11 million galaxies (11,171,958) at a median redshift of about 0.4.  This is by far the most galaxies used to conduct this measurement, and our depth is much greater than that of the recent 2MASS and 2dRGRS measurements.  This base data set was split into five sub-samples, three with magnitude ranges $18 \leq r <  19$, $19 \leq r <  20$,  and $20 \leq r <  21$, a sample of early-type ($u-r > 2.2$) galaxies, and a sample of late-type ($u-r \leq 2.2$) galaxies.  The bulk of our analysis centers on this base data set and its five subsamples.

\section{Methodology}
\subsection{Angular Correlation functions}
We estimate N-point area-averaged angular correlation functions, $\bar{\omega}_{N}$($\theta$), using a counts-in-cells technique (e.g., \citealt{P80,Gaz94, Sza02}).  We divide DR3 into cells of equal angular area, so that each cell $i$ contains some number of galaxies ($n_{i}$).  The average number of galaxies in a cell is then $\bar{n}$.  The over density for cell $i$ is thus

 \begin{equation}
 \delta_{i} = \frac{\bar{n} - n_{i}}{\bar{n}}
 \end{equation}
 
\noindent The central moment of the angular counts, $m_{N}(\theta)$, is calculated over all of the cells

\begin{equation}
 m_{N}(\theta) = \frac{1}{n_{c}(\theta)}\sum_{0}^{n_{c}(\theta)}{(\bar{n} - n_{i})^N(\theta)} = \bar{n}^N(\theta)\langle\delta_{i}^N(\theta)\rangle,
 \end{equation}

\noindent where $\delta_{i}^N(\theta)$ and $\bar{n}(\theta)$, are for cells with an area that has an equivalent circular radius to the angular scale ($\theta$) being investigated, $n_{c}(\theta)$ is the total number of cells at that angular scale, and $\langle...\rangle$ represents a statistical average.  The area-averaged N-point functions, $\bar{\omega}_{N}$($\theta$), can be explicitly calculated from $m_{N}(\theta)$ via the connected central moments, $\mu_{N}(\theta) = \bar{n}^N(\theta)\langle\delta_{i}^N(\theta)\rangle_{c}$, as shown (for $N < 7$) in Appendix~\ref{app:a}.  The subscript $c$ in $\langle\delta_{i}^N(\theta)\rangle_{c}$ refers to the connected part of the expectation value, and we include it because for $N > 2$, $\langle\delta_{i}^N(\theta)\rangle_{c}$ $\neq$ $\langle\delta_{i}^N(\theta)\rangle$ but requires the subtraction of factors of lower order moments.  The connected moment would equal $\bar{\omega}_{N}$($\theta$) if not for shot-noise error, $\delta_{SN}$, due to the discreteness of the counts-in-cells statistic, i.e.

 \begin{equation}
 \bar{\omega}_{N}(\theta) = \langle\delta_{i}^N(\theta)\rangle_{c} - \delta_{SN} 
 \end{equation}
  
 % NOTE FOR ASHLEY: you can insert correct references, but presumably \citealt{Gro77,Sza92,Gaz94} are OK for now
 
In a hierarchical model (e.g., \citealt{Gro77,Sza92,Gaz94}), higher-order correlations can be expressed in terms of the two-point correlation function and the volume-averaged correlations are given by

\begin{equation}
\bar{\xi}_{N}(R) = S_{N}[\bar{\xi}_{2}(R)]^{N-1}
\end{equation}

\noindent where $S_{N}$ is the hierarchical amplitude.  In a similar manner, we can define the analogous relationship for the area-averaged angular correlations

 \begin{equation}
s_{N} \equiv  \frac{\bar{\omega}_{N}(\theta)}{[\bar{\omega}_{2}(\theta)]^{N-1}}
\end{equation}

The hierarchical amplitudes of the higher-order moments encode much of the pertinent information on the distribution of the data. These amplitudes, therefore, embody the central analysis of this paper.

Cross-correlations between the galaxy density and contaminants that could add systematic errors to our measurement are also calculated.  As in \citet{Scr02}, the cross correlation at an angular bin $\theta_{\alpha}$ is given by

\begin{equation}
\omega_{gc}(\theta_{\alpha}) = \frac{1}{N(\theta_{\alpha})}{\sum_{i,j}^{}{\delta_{i}^{g}\delta_{j}^{c}\theta_{ij}^{\alpha}}}
\end{equation}

\noindent where $\theta_{ij}^{\alpha}$ is unity if the separation between cells $i$ and $j$ is within the angular bin $\theta_{\alpha}$ and zero otherwise, subscript $c$ denotes a particular contaminant (e.g., reddening, seeing), and $N(\theta_{\alpha})$ is just the number of cells within the angular bin $\theta_{\alpha}$.

\subsection{Inversion to Three Dimensions}
To compare results of angular correlation measurements to the predictions of perturbation theory and to constrain galaxy bias, it is easiest to deproject the two dimensional measurement into a three dimensional measurement.  Given the redshift distribution of the source population, the angular hierarchical amplitude ($s_{N}$) can be converted directly to an estimate of the real-space amplitude ($S_{N}$).  The relationship is \citep{Gaz94}, 
\begin{equation}
S_{N}(\bar{r}) = \frac{s_{N}(\theta)B_{N}(\gamma)}{r_{N}(\gamma)C_{N}(\gamma)},
\label{eq:DP}
\end{equation}
where $\bar{r}=2\theta$ is the mean scale probed for a survey of given depth $D$ at angular scale $\theta$ ($\theta \ll 1$ radian), and $\gamma$ is the slope of the real-space correlation function, which we determine to be 1.75 (see \S 5.1).  The values $B_{N}(\gamma)$ and $C_{N}(\gamma)$ were determined by \cite{Gaz94} by considering the number of different configurations of hierarchical tree-graphs.  They are of order unity, but non-negligible for large $N$.  The remaining factor ($r_{N}$) is determined by the selection function ($\psi$) for the sample of objects:
\begin{equation}
r_{N} =\frac{I_{1}^{N-2}I_{N}}{I_{2}^{N-1}},
\end{equation}
\begin{equation}
I_{N}=\int_{0}^{\infty}\psi^{N}x^{(3-\gamma)(N-1)}(1+z)^{(3+\epsilon-\gamma)(1-N)}F(x)x^2dx,
\end{equation}
where $x$ is the comoving distance, $\epsilon$ is a factor that describes the evolution of clustering, which we take to be zero, and $F(x)$ is a curvature term that is unity for a flat universe.  

When $\frac{dn}{dz}$ is known for the projected sample, equation 9 is readily evaluated in redshift space.  Using the transformations $\frac{dn}{dx} = x^2\psi$, $(\frac{dn}{dx})^{N}dx=(\frac{dn}{dz})^{N}(\frac{dz}{dx})^{N-1}dz$, and $\frac{dz}{dx} = \frac{H(z)}{c}$, where $n$ is the number of objects, $c$ the speed of light, and $H(z)$ the Hubble parameter, yields
\begin{equation}
I_{N}=\int_{0}^{\infty}x^{(1-\gamma)(N-1)}\left(\frac{H(z)}{c}\right)^{N-1}(1+z)^{(3-\gamma)(1-N)}\left(\frac{dn}{dz}\right)^{N}dz.
\end{equation}

We will also wish to determine the relationship between $\bar{\omega}_{2}$($\theta$) and $\bar{\xi}_{2}(R)$.  Assuming the forms \citep{P80}:
\begin{equation}
\bar{\xi}_{2}(R,z)=\left(\frac{R}{R_{o}}\right)^{-\gamma}(1+z)^{-(3+\epsilon)}
\label{eq:real2pt}
\end{equation}
and
\begin{equation}
\bar{\omega}_{2}(\theta)=A\theta^{1-\gamma}
\label{eq:real2pt2}
\end{equation}
the constant $A$ is given by
\begin{equation}
A = H_{\gamma}I_{2}R_{o}^{\gamma}
\label{eq:real2pt3}
\end{equation}
where $H_{\gamma} = \Gamma(0.5)\Gamma(0.5\gamma - 0.5)/\Gamma(0.5\gamma)$, $\Gamma$ is the gamma function.  Given both $A$ and $\gamma -1$, we can derive $R_{o}$ and thus the form of $\bar{\xi}_{2}(R,z)$.

In order to determine $\frac{dn}{dz}$, we used the photometric redshifts of galaxies from the SDSS first-data release (DR1; \citealt{Ab01}).  We construct $\frac{dn}{dz}$ by using each published redshift and its error (rejecting any with error greater than twenty percent) to create a probability density function (PDF).  The PDFs for each redshift are combined to produce the expected number of objects ($n$) in a redshift bin $\Delta z = 10^{-3}$.  This distribution can then be normalized and interpolated over in order to estimate $\frac{dn}{dz}$.  For SDSS galaxies with $18 \leq r < 21$, the resulting normalized $n(z)$ is plotted in Figure 1.  The distribution of $n(z)$ is smooth and roughly Gaussian.  Since DR1 is a sub-sample of DR3, the redshift selection functions are nearly identical.  The $n(z)$ plotted in Figure 1 is thus an acceptable estimate of the true $n(z)$ for our full sample of galaxies.  For our five subsamples (see \S 2), we likewise use the corresponding DR1 photometric redshifts to obtain estimates of their redshift selection functions.

\subsection{Pixelation}
\label{sec:pixel}
% NOTE FOR ASHLEY: ADM rewrote a lot of this as it was in the passive voice.

Tegmark, Xu, and Scranton\footnote{http://lahmu.phyast.pitt.edu/$\sim$scranton/SDSSPix/}, have developed a software package (henceforth \SDSSpixb) that pixelizes the sky in a manner convenient for SDSS data.  Working in SDSS $\lambda$/$\eta$ coordinates (see \citealt{Sto02}), \SDSSpixt creates almost-equal-area, pseudo-rectangular pixels over the entire sky.  \SDSSpixt breaks SDSS stripes into a number of rectangles (transcribed on the surface of the sphere), and hierarchically bisects parent rectangles into increasingly smaller rectangles (pixels).  Thus, to probe any desired angular scale, one only needs to bisect the parent \SDSSpixt rectangles down to the corresponding resolution in the \SDSSpixt equal-area schema. Thus, given information for pixels at the smallest angular scale (the base resolution), gaining information at larger angular scales is trivial.  Statistical counts can be easily, and rapidly, calculated for large angles simply by accurately determining them for the smallest scale of interest.

For this work, the basic methods of \SDSSpixt were utilized for the selected pixelation schemes.  \SDSSpixt pixelates the entire sky, which leads to a burdensome number of pixels at small scales.  Therefore, at small scales ($\theta \lesssim$ 0.1 degrees) we reimplemented the algorithms used to make equal area pixels over the entire sky, in a manner that took advantage of the survey strategy.  DR3 is broken into 18 {\it stripes}.  If one uses the survey $\lambda$/$\eta$ coordinates, each stripe makes a rectangle that is 2.5 degrees wide in the $\eta$ coordinate and typically between 90 and 110 degrees in the $\lambda$ coordinate.  Since each stripe has a unique range in $\lambda$, the SDSS DR3 essentially consists of 18 data subsections.  Each stripe was thus given its own set of pixels based on its $\lambda$ coordinate, meaning that the majority of the sky, not covered by DR3, could quickly be discarded.  The correlations were measured for each stripe, with the full measurement being the composite average over all stripes, inverse-variance-weighted by the number of objects in that stripe.  

For larger angular scales ($\theta \gtrsim$ 0.1 degrees), averages over the entire sample are required to obtain an accurate result.  On these scales, pixelating the entire sky is less computationally challenging, and \SDSSpixt was directly used for these scales.  As a result, the correlations were measured at these scales over the entire DR3 area.  We have verified that the results using the striped method are consistent with using the default \SDSSpixt implementation at the scale of 0.1 degrees.  From here on, we will refer to the method used for small angular scales as the {\it{striped}} method and to the method for large angular scales as the \SDSSpixt method.

\subsection{P(n) plots}
For each resolution we measured the probability of finding a cell with $n$ counts in it, $P(n)$.  This serves to justify the pixelization scheme used.  Ideally, these plots show a smooth curve centered at the most probable count, as this signifies sufficient sampling of the data.  Figure 2 shows the $P(n)$ plots for resolutions that correspond to angular scales between 0.02 and 5.2 degrees.  The $P(n)$ become less smooth as the angular scale gets larger but appear sufficient for the purposes of our measurements to an angular scale of about 1.3$^0$.  At a scale of 2.6$^0$, Figure 2 suggests that a measurement will have a large associated statistical error.  At a scale of 5.2$^0$, it appears unlikely that any information can be determined to any precision.  This is due to the fact that we use non-overlapping cells in our analysis and thus have few pixels at this large of a scale.  As a result, we expect to be able to make precision measurements to an angular scale of $\sim 1^0$, and relevant measurements to an angular scale of $\sim 3^0$. 

\subsection{Errors and Covariance}
\label{sec:err}

We compute errors and covariance matrix using a jackknife method (see, e.g., \citealt{Scr02,Mye03,Mye05a}). To properly constrain fit parameters, we calculate covariance matrices by following the prescription of~\cite{Mye06}, which entails splitting the data into $N$ subsamples and calculating the covariance between angular bins $\theta_i$ and $\theta_j$  as

\begin{equation}
C_{i,j }= C(\theta_i,\theta_j) = \frac{\sum_{k=1}^{N}[\bar{\omega}_{full}(\theta_i) - \bar{\omega}_{k}(\theta_i)][\bar{\omega}_{full}(\theta_j) - \bar{\omega}_{k}(\theta_j)]n_{gal}^{2}}{\sum_{i=1}^{N}n_{gal}^{2}}
\label{eq:JK}
\end{equation} 

\noindent where $\bar{\omega}_{k}(\theta)$ is the value for the correlation measurement omitting the $k$th subsample of data, and the number of galaxies in subsample $N$, $n_{gal}$, is used to inverse-variance weight the contribution of that subsample to the covariance.  The jackknife errors, $\sigma_i$ can be obtained from the diagonal elements of the covariance matrix

\begin{equation}
\sigma_{i}^{2} =  C_{i,i}
\label{eq:JK2}
\end{equation}

\noindent This jackknife technique is used to calculate the covariance for every order correlation.  We estimate $\chi^{2}$ fits to model functions ($\bar{\omega}_{m}$) using the inverse of the covariance matrix.

\begin{equation}
\chi^2 = \sum_{i,j}[\bar{\omega}(\theta_i) - \bar{\omega}_{m}(\theta_i)]C_{i,j}^{-1}[\bar{\omega}(\theta_j) - \bar{\omega}_{m}(\theta_j)]
\label{eq:Chi}
\end{equation}

\noindent All of the averages and power-law fits conducted throughout the rest of the paper are accomplished by minimizing the $\chi^2$ defined in Equation \ref{eq:Chi}.

For the {\it{striped}} method, we utilize the natural geometry of the SDSS.  Each of the 18 different stripes in the DR3 forms a natural subset of the overall data.  The the covariance matrix is thus calculated using each of the possible subsamples of DR3 that is made up of 17 stripes.  For the angular scales that we utilize the {\it{striped}} method, we find that these 18 subsamples are sufficient to create a stable covariance matrix.  For the larger angular scales probed by the \SDSSpixt method, we find that many more jackknife subsamples are necessary.  Our covariance matrix is not stable until we utilize at least 100 jackknife subsamples.  We therefore use 150 jackknife subsamples for all calculations involving the \SDSSpixt method.

\section{Masks}

We will generally refer to useful observational information (such as seeing and Galactic extinction values) across each pixel in our schema (see \S\ref{sec:pixel}) as forming a mask of that information. The DR3 area required significant masking to be statistically useful.  For instance, any pixels at the base resolution of our schema that intersect the standard SDSS imaging masks (which cover areas missing from DR3 due to, for example, satellite trails, charge transfer, bright objects\footnote{http://www.sdss.org/dr3/products/images/use$\_$masks.html}) are discarded (masked) from our analyses.  During our tests of stripe variation, we found that a mask for galaxy M101 ($\alpha$ = 14$^{h}$ 03$^{m}$ 12$^{s}$, $\delta$ = 54 21\arcmin~$\!$$\!$ 00\arcsec) was not included in the standard DR3 imaging masks, and thus we also mask an area of $\sim$ 0.25 square degrees around the center of M101. We further discard pixels based on their mean value of seeing or reddening, as will be described in the the subsequent sections.

Pixels are masked out at the base resolution of the pixelization scheme (the smallest angular scale that we calculate, see \S\ref{sec:pixel}).  For pixels above the base resolution, the fraction of the area of the pixel that is unmasked is calculated using the base pixels.  The over density (previously defined by Equation 1 in \S 3), $\delta_{i}$ in pixel $i$ becomes 
\begin{equation}
 \delta_{i} = \frac{\bar{n} - \frac{n_{i}}{\Delta_{i}}}{\bar{n}}
 \end{equation}

\noindent where $\Delta_{i}$ is the fractional area of cell $i$.  This allows for a complete mask that is consistent on all angular scales.

\subsection{Reddening}

As shown by \citet{Scr02}, including areas of high Galactic obscuration (reddening), can alter statistical counts for extragalactic objects.  We therefore tested to see how this Galactic reddening affected our analyses.  We concentrated our analysis on the large angle measurements since it has been shown (e.g., \citealt{Scr02}, \citealt{Myers06}) that reddening affects correlation measurements on scales larger than $\sim$ 0.1 degrees.  In order to determine how reddening affects the eventual hierarchical amplitudes, we first tested the result of the measurement of $s_{3}$ for galaxies with $18 \leq r < 21$ and $g$-band reddening cuts at $A_{g} <$ 0.35, 0.3, 0.25, 0.2, and 0.15, as shown in Figure 3.  At any angular scale, the effect of changing the reddening cut appears to be negligible.  Figure 4 shows the relationship between the measured $s_{3}$ at 1.3$^0$, and the reddening cut, between $A_{g} <$ 0.5 and  $A_{g} <$ 0.175.  There is a slight trend for the amplitudes to decrease, but this is negligible given the size of the error bars.  Based on the SDSS EDR, \citet{Scr02} determine a reddening cut of $A_{r} <$ 0.2 ($\sim A_{g}< 0.3$) to be appropriate.    In Figures 3 and 4, the values at this reddening cut have small errors compared to both less and more restrictive cuts.  Thus, we also use a reddening cut at $A_{g} <$ 0.3 throughout our analyses. 

\subsection{Seeing}

We have also tested the effects of seeing in our correlation measurements.  The hierarchical amplitude $s_{3}$ is shown in Figure~\ref{fig:5} for seeing cuts at 1\farcs4, 1\farcs5, 1\farcs6, 1\farcs7, 1\farcs9, and 2\farcs2, for galaxies with $18 \leq r < 21$.  There are noticeable differences in the value of $s_{3}$ depending on the seeing cut.  The error on the measurement appears to be minimized at a seeing cut of 1\farcs5 for the majority of the angular scales. 

To further quantify the effect of seeing in the correlation measurements, the cross-correlation between galaxy density and the mean seeing per pixel was determined for galaxies with $18 \leq r < 21$ for seeing cuts between 1\farcs9 and 1\farcs4.  This data is presented in Figure 6.  The ideal seeing cut is one that produces a cross-correlation that is consistent with being flat and with zero.  None of the seeing cuts produce results that are flat, but the seeing cut at 1\farcs5 produces a correlation that is most consistent with zero at large scales.  The fact that the seeing-galaxy cross correlation increases at small scales should not be a problem, because this increase is insignificant in comparison to the amplitude of the galaxy auto-correlation at these scales.  Thus, a seeing cut at 1\farcs5 is used throughout our analyses.

\subsection{Variation Between Stripes}

As a final systematic test, we looked at the variation in our correlation measurements as a function of the SDSS stripe.  Some variation between stripes was expected, but any stripes whose values deviated seriously from the mean could signify a problem with the data in that stripe.  Figure~\ref{fig:7} shows the measured $\bar{\omega}_{2}(\theta)$ and $s_{3}(\theta)$ for galaxies with $18 \leq r < 21$, measured in each individual stripe used for this work.  There are no significant differences between stripes, and thus no one stripe significantly alters the outcome of our measurements.  We repeated this analysis for each of our five subsamples (see \S 2), and found that no one stripe significantly altered the outcome of any measurements.  As a result, we use all 18 stripes in our analysis.  We note that this would not have been the case had we not masked the area around M101 from our analysis (see above).
\section{Results}

\subsection{Area-averaged Correlation Functions and Hierarchical Amplitudes}

Figure 8 shows the area-averaged correlation functions for $N \leq 7$ determined for galaxies with $18 \leq r < 21$.  Errors for each point were determined by the jackknife method (Equations \ref{eq:JK} and  \ref{eq:JK2}).  For each $N$, the correlation function has a shape consistent with a power-law, but for $N > 2$ errors dominate at scales large than $\sim$ $1^o$.  Figure 8 displays the result using both the striped and the \SDSSpixt method between about 0.08 and 0.4 degrees showing the two methods are in agreement.  Assuming a form $\bar{\omega}_{2} = A\theta^{1-\gamma}$, the $\chi^2$ best model fit for $N = 2$ over the angular range $0.1<\theta < 2^o$ is $A =  (7.45 \pm 0.02)\times10^{-3}$ and $\gamma = 1.757 \pm 0.002$.  These results are consistent with previous measurements (see, e.g., \citealt{Gaz94,Con02, FR05}).  

The hierarchical amplitudes $s_{N}(\theta)$ for $N \leq 7$ measured in magnitude ranges $18 \leq r < 19$,  $19 \leq r < 20$, $20 \leq r<21$, and $18 \leq r < 21$ are shown in Figure~\ref{fig:9}.  The errors on the measurement of $s_{N}(\theta)$ are calculated using Equations \ref{eq:JK} and \ref{eq:JK2}.  Our results show that the amplitudes remain roughly constant over this angular range, although there is interesting structure.  The differences between the magnitude ranges are more easily compared if one deprojects the $s_N$ to obtain estimates of the real-space amplitudes, $S_{N}$ (see Equation \ref{eq:DP}).  These amplitudes, plotted over the same magnitude ranges as the $s_{N}$, are displayed in Figure 10.  For each magnitude range, the shape of $S_{3}$ is remarkably similar.  Its value is nearly constant until $R$ reaches a scale of about 10 $h^{-1}$ Mpc after which it declines sharply.  

For $N > 3$, the shape is not as consistent between the different magnitude bins, but strong similarities remain.  Interestingly, the brightest magnitude bin has the most consistently constant amplitudes and the error bars are not significantly larger than those in the other magnitude bins (and in some cases, they are significantly smaller), despite having the fewest number of objects ($\sim$ 1 million galaxies in $18 \leq r < 19$; $\sim$ 2.75 million in $19 \leq r < 20$; and $\sim$ 7.5 million in $20 \leq r<21$).  This is likely due to the fact that the brightest magnitude bin has the smallest spread in redshift and is therefore sampling the smallest volume.  All four magnitude bins show a dramatic dip in the amplitudes between 1 and 5 $h^{-1}$ Mpc, with subsequent flattening or increase in the calculated amplitude values.  This feature is amplified with increasing $N$ such that for $N = 7$ the decrease is nearly an order of magnitude, while for $N = 4$ it is closer to a 50$\%$ decrease.  The fact that we see the same general features at the same physical scales (and thus different angular scales) suggests that the features we see are real and are not caused by our treatment of the data.

We further quantify the degree to which the amplitudes are constant by assuming a relationship $S_{N} = B_{N}*\theta^{\alpha_{N}}$ and comparing $\alpha_{N}$ to zero.  Splitting the data for the magnitude range $18 < r < 21$ into regions $0.2 < \bar{r} < 2.0$ and $2.0 < \bar{r} < 10$ $h^{-1}$ Mpc and performing $\chi^2$ fits to the assumed power-law behavior gives the results displayed in Table \ref{T:slope1}.  These results are not consistent with zero slope.  Repeating for galaxies in the magnitude range $18 \leq r < 19$, one obtains the results in Table \ref{T:slope2}. The brighter galaxies have amplitudes that are more consistent with constant values.  Although for both samples the general nature of the amplitudes is a hierarchy of roughly constant $S_{N}$, there remain clear differences in the measured amplitudes between bright galaxies and the full sample.  
 
It is instructive to find the average values of the hierarchical amplitudes for comparison with previous studies and the predictions of Perturbation Theory.  We perform $\chi^2$ fits to a constant value for both $s_{N}$ and $S_{N}$ for $N \leq 7$ between 0.2 and 10 $h^{-1}$ Mpc in the $18 \leq r < 19$ and $18 \leq r <21$ magnitude ranges.  These results are presented in Table 3.  The values of $s_{N}$ are in fair agreement with previous results determined for the APM Survey \citep{Gaz94} and for the 2MASS extended source catalog \citep{FR05}, though the data taken from $18 \leq r < 19$ are in better agreement, especially at lower order.  One would not expect a perfect match between the results of different surveys given that it is clear that the amplitudes are not constant, different surveys will select different spectral types, and the different techniques were used to estimate the redshift distributions. 

\subsection{N-point Correlations for Early and Late-type Galaxies\label{sec:npoint}} 
\cite{Stra01} provide simple color criteria that can be used to separate early- ($u-r > 2.2$) and late-type ($u-r \leq 2.2$) galaxies from SDSS data.  Using this color cut, we separate our sample into early- and late-type galaxies and repeat our N-point measurements for these two samples.  It is important to note that the redshift distributions for early- and late-type galaxies in the magnitude range $18 \leq r < 21$ are quite similar.  The median redshift of the early-type galaxies is 0.361, while it is 0.374 for late-type.  The full-width-half-maximums of the distributions are 0.35 and 0.39, respectively.  It is therefore reasonable to expect that the two populations primarily sample the same epochs.  Thus, we determine that direct comparison of the correlation functions determined for each galaxy type is fair.

 Figure 11 shows the results of the area-averaged correlation measurements for early- and late-type galaxies.  The early-type galaxies clearly show stronger clustering at all angular scales, in agreement with previous results (e.g., \citealt{W98,Z02,N02,Ma03}), and their correlation functions display a power-law behavior for all $N$ to large angular scales.  The $\chi^2$ best model fit for $1-\gamma$ over scales $0.01<\theta < 2^o$ for early-types is -0.846 $\pm$ 0.002.  The late-type galaxies, on the other hand, look quantitatively different.  The magnitude of the slope of the correlation functions appears to decrease between $\theta \sim 0.01$ and $\theta \sim 1.0$, with the transition occurring around $\theta \sim 0.1$. The difference in slope at small angles relative to larger angles becomes even more pronounced for the high-order angular correlations.  Determined over the same angular scales as the early-types, the $\chi^2$ best model fit for $1-\gamma$ for late-types is -0.677 $\pm$ 0.002---significantly more shallow than that for early-types.  
 
The transition between the small-angle and large-angle slopes of the late-type $\bar{\omega}_N$ results is smooth rather than abrupt, and actually appears as an inflection point at scales corresponding to about 1 $h^{-1}$ Mpc, similar to the inflection point observed by \cite{Z05}.  This inflection point is more clearly pronounced for late-type galaxies, and the transition region itself becomes more pronounced as $N$ increases.  In addition, the small-angle slope of the late-type galaxy correlation functions increases as the order of the correlation function itself increases.  Conversely, the early-type galaxies do not show any dramatic change with increasing $N$.  We discuss these observations in more detail in section~\ref{sec:EandL}.

The hierarchical amplitudes $s_{N}$ and $S_{N}$ for early- and late-type galaxies with $18 \leq r < 21$ are plotted in Figure 12.  Interestingly, the error on individual points is larger on the measurement of the late-type correlation functions, despite the fact that we find more late-type galaxies (6,223,510) than early-type galaxies (4,948,448).  This might imply greater cosmic variance in the clustering of those objects we classify as ``late-type".  

The early-type measurements show behavior that is quite similar to the measurement for all galaxies, (over $2 < \bar{r} < 10$ $h^{-1}$ Mpc, $\alpha_{3} = -0.23 \pm 0.04$ and $\alpha_{5} = -1.1 \pm 0.1$; over $0.2 < \bar{r} < 2.0$ $h^{-1}$ Mpc, $\alpha_{3} = 0.07 \pm 0.02$, and $\alpha_{5} = 0.41 \pm 0.08$).  These values are quite close to those determined using all galaxy types, despite the fact that there is a large difference in the slopes of $\bar{\omega}_{N,early}$ and $\bar{\omega}_{N,all}$.  At  scales $\bar{r} > 2$ $h^{-1}$ Mpc the $S_{N}$ for late-type galaxies are marginally consistent with being constant (e.g., $\alpha_{3} = -0.12 \pm 0.07$).  At small scales, however ($\bar{r} < 2$ $h^{-1}$ Mpc), the $S_{N}$ are highly inconsistent with being constant (e.g., $\alpha_{3} = -0.33 \pm 0.04$, $\alpha_{5} = -4.0 \pm 0.3$).  Table 4, respectively, present the $\chi^2$ best fit average values of $s_{N}$ and $S_{N}$ for early- and late-type galaxies in the range $1 < \bar{r} < 10$ $h^{-1}$ Mpc.  The average values of $s_{N}$ and $S_{N}$ for early-type galaxies are significantly higher than those for late-types, and the factor increases through order $N = 7$. 

Figure \ref{fig:ratio} shows the ratios of $S_{3}$ and $S_{4}$ for early- to late-type galaxies.  Both ratios show significant structure.  There is a maximum in both ratios at $\sim$ 1 $h^{-1}$ Mpc.  At the smallest scales, $S_{N,late}$ is larger than $S_{N,early}$.  Figure 12 shows that the small scale structure is largely dependent on the nature of $S_{N,late}$.  Figure 13 shows that at scales larger than $\sim$ 1 $h^{-1}$ Mpc, the ratios are roughly constant.  The ratios of $S_{3}$ are slightly less than two and the ratios of $S_{4}$ are close to three at these scales.  This startling difference between the grouping behavior of early- and late-type galaxies as a function of scale can be speculatively explained by several scenarios, which we will discuss in some depth in section~\ref{sec:EandL}. In general, our results trace higher-order biasing schema, and can be well described by differences in bias between galaxy types.  We explore biasing constraints in the following section.  

\section{Discussion}
Our general results are consistent with previous measurements (e.g., \citealt{Gaz94,Sza02,Cr04,FR05}).  The measured $S_{N}$ display the hierarchical scaling that one expects to observe for an initially Gaussian density field that has evolved due to gravitational instabilities.  To properly constrain primordial Gaussianity, we would need significant measurements at scales where there is linear clustering ($\bar{r} \gtrsim 10~h^{-1}$ Mpc).  Since this is where errors begin to dominate our measurement, we do not place any strong constraints on primordial Gaussianity.  We note that there are no structures in the $S_{N}$ measurements at scales greater than 4 $h^{-1}$ Mpc that are consistent over all magnitude bins and all $N$.  The $S_{N}$ are, however, consistent with constant amplitudes at the 2$\sigma$ level for all $N$ and all magnitude bins, consistent with Gaussian initial conditions. 

\subsection{First-Order Galaxy Bias}
Galaxies are tracers of the underlying dark matter distribution.  As a result there is no guarantee that galaxies cluster with the same amplitude as the underlying dark matter.  At scales where the clustering is linear or quasi-linear, one can relate the over-density, $\delta_{DM}$, of dark matter halos to the measured over-density of galaxies, $\delta_{g}$, by \citep{Fry93}:
\begin{equation}
\delta_{g} = \sum_{k=0}^{\infty}{\frac{b_{k}}{k!}\delta_{DM}}
\label{eq:bias}
\end{equation} 
The first-order bias is related to the measured $\bar{\xi}_{2}$ and the dark matter $\bar{\xi}_{2}$ by
\begin{equation}
\bar{\xi}_{2,T} = b_{1,T}^{2}\bar{\xi}_{2,DM}
\end{equation}
where the subscript $T$ stands for the galaxy sample (either {\it early, late,} or {\it all}; all refers to magnitude bin $18 \leq r < 21$ unless otherwise noted) used in the measurement.  By using Equations \ref{eq:real2pt}-\ref{eq:real2pt3}, we can determine the relative first order bias ($b_{1}$) for each galaxy sample: 
\begin{equation}
b_{1,T} = \left(\frac{R_{o,T}}{R_{o,DM}}\right)^{\gamma/2}
\end{equation}
Galaxies, as a whole, have negligible first-order bias (e.g., \citealt{Sel05,Gaz05,Pan05}).  We thus take $R_{o,DM} = R_{o,all}$ and thereby determine $b_{1}$ for early- and late-type galaxies.  We find that $b_{1,early} = 1.36 \pm 0.04$ and $b_{1,late} = 0.81 \pm 0.01$.  This is consistent with findings that early-type galaxies cluster more strongly than late-types (e.g., \citealt{W98,Z02,N02,Ma03}), but the ratio of early- to late-type bias of $\sim 1.75$ is significantly larger than that found by \cite{W98} of $\sim 1.2$.  The ratio of $b_{1,early}$ to $b_{1,late}$ is largely independent of the assumption that $b_{1,all} = 1$ (a 50$\%$ change in $R_{o,DM}$ results in a 6$\%$ change in the ratio). 

Assuming that bias terms $b_{k}$ are negligible for $k \geq 2$, a simple relationship follows from Equation \ref{eq:bias} that $S_{N,T} = b_{1,T}^{2-N}S_{N,DM}$.  Using this relationship and the average values of $S_{N}$ for $R$ between 1 and 10 $h^{-1}$ Mpc (where both $S_{N,early}$ and $S_{N,late}$ have nearly zero slope) and $N = 3,4,5$, we calculate the ratio between $b_{1,early}$ and $b_{1,late}$.  The results for $N = 3,4,5$ are, respectively, $0.74 \pm 0.11$,  $0.73 \pm 0.22$, and $0.50 \pm 2.3$.  These are quite consistent with each other, but wildly inconsistent with both the results found from $\bar{\omega}_{2}$ and the notion that early-type galaxies cluster more strongly than late-type galaxies (e.g., \citealt{W98,Z02,N02,Ma03}).  Clearly, assuming higher-order bias terms are negligible leads to a startling contradiction.  This inconsistency can be resolved only by incorporating higher order bias terms.

\subsection{Higher-Order Galaxy Bias}
Using Equation \ref{eq:bias} a relationship can derived, valid on linear and quasi-linear scales, between the measured $S_{N,T}$, the $S_{N,DM}$, $b_{1,T}$, and $c_{N,T}$ ($c_{N} = \frac{b_{N}}{b_{1}}$).  For $N = 3$, this is given by \citep{Fry93}
\begin{equation}
S_{3,T}=b_{1,T}^{-1}(S_{3,DM} + 3c_{2,T})
\end{equation}
and for $N = 4$,
\begin{equation}
S_{4,T}=b_{1,T}^{-2}(S_{4,DM} + 12c_{2,T}S_{3,DM}+4c_{3,T}+12c_{2,T}^{2})
\end{equation}
Using perturbation theory \cite{J93} determined $S_{3,DM}$ to be $\frac{34}{7} - (n+3)$ by , where $n$ is the slope of the matter power spectrum.  Recently, several groups have converged to a canonical value of $n =-2$ to within 5$\%$ (e.g., \citealt{Per01,Teg02,Cole05}); thus, we use $n = -2$ for our analysis.  Since Equations 21 and 22 are valid only in the quasi-linear and linear regimes, and since error dominates our measurement at scales greater than 10 $h^{-1}$ Mpc, we calculate the $\chi^2$ best fit average value of $c_{2}$ in the range between 4 and 10 $h^{-1}$ Mpc.  We find $c_{2,all} = -0.24 \pm 0.08$, $c_{2,early} = 0.28 \pm 0.10$, and $c_{2,late} = -0.70 \pm 0.08$.  As a consistency check, we also measure $c_{2,all}$ for galaxies in the magnitude bin $18 \leq r < 19$, finding $c_{2,all} = -0.31 \pm 0.08$.  The two measurements are consistent.

The result for all galaxies is entirely consistent with the result of \cite{Gaz05} of $c_{2}=-0.36_{-0.09}^{+0.13}$.  If we are to use the value of $b_{1}$ determined by \cite{Gaz05} of 0.94$_{-0.11}^{+0.13}$, the resulting $c_{2}$ is $-0.30 \pm 0.15$, which is in remarkable agreement with their previous measurement.  Our $c_{2}$ is slightly inconsistent with the results of \cite{Pan05} and \cite{H05}, but as noted by \cite{Gaz05}, differences between measurements should not be surprising given that they are conducted in different ways and/or use different galaxy catalogs.  In fact, given our results that early- and late-type galaxies have different second order bias properties, measurements using different galaxy catalogs will likely result in different $c_{2}$ measurements unless the fractions of early- and late-type galaxies are nearly identical in the two samples.

The positive result for $c_{2,early}$ is consistent with the recent measurement of $c_{2} = 0.57 \pm 0.33$ made by \cite{FR05} and supports their assertion that their measurement of positive second order bias was likely due to the dominance of early-type galaxies in their 2MASS sample.  The second-order bias we measure for late-type galaxies is significantly negative, and it is inconsistent with zero at $9\sigma$.  

Given the values of $c_{2}$, one can go on to determine the significance of the third-order bias.  \cite{Ber94} finds that for $n = -2$, $S_{DM,4} = \frac{36457}{1323}$,  which results in $c_{3,all} = 0.98 \pm 0.89$, $c_{3,early} = 1.46 \pm 1.37$ and $c_{3,late} = 0.63 \pm 1.04$.  These are all positive, though all are consistent with zero at about the 1$\sigma$ level.  It seems likely, based on these results, that given extremely precise measurements of $S_{N}$ and analytic expressions for $S_{DM,N}$, that $c_{N}$ would be found to be non-neglibible for all $N$.  Table \ref{T:bias} summarizes all of our bias measurement results.

\subsection{Early- and Late-type Galaxies}
\label{sec:EandL}
In the previous section, we quantified the differences between the clustering behavior of our early-type and late-type galaxy samples. Figures~\ref{fig:12} and~\ref{fig:ratio}, however, also can provide qualitative insights into the clustering differences between these two populations. In this section, we explore these two figures in more detail, focusing on the interpretation of Figure~\ref{fig:12} in light of the halo occupation distribution framework~\citep[HOD; see, e.g.,][]{Berl02}, and on the interpretation of Figure~\ref{fig:ratio} in terms of the interplay between star formation and local density.

In Figure~\ref{fig:12}, we observe several important effects, previously detailed in Section~\ref{sec:npoint}. Specifically, these are that (1) early-type galaxies always cluster more strongly than late-type galaxies; (2) the slope of the late-type galaxy correlation function increases on smaller angular scales; (3) this change in slope is smooth, indicating an inflection point; and (4) both the small-angle slope and the width of the transition between the large-angle and small-angle late-type galaxy correlation functions increase for higher orders. Of these effects, the last three provide insight into the distribution of different galaxies within dark matter halos.

The shape of $\bar{\omega}_2$ for our early- and late-type galaxy samples appears to be consistent with the HOD determined by \cite{Z05} of red (early-type) galaxies and blue (late-type)  galaxies. \cite{Z05} find that smallest mass halos ($\lesssim 10^{13} h^{-1}$ M$_{\odot}$) are predominantly occupied by late-type galaxies, while larger mass halos ($\sim 3 \times 10^{14} h^{-1}$ M$_{\odot}$) contain the majority of the early-type galaxies and also a large fraction of the late-type galaxies.  Thus, to model the late-type galaxy correlation function, a combination of one-halo and two-halo terms must be used, which naturally results a smooth transition between different correlation function slopes at small angles and large angles. 

We note, however, that our work differs from the \cite{Z05} work in two ways, which is reflected in Figure~\ref{fig:12}. First, by using a spectroscopic sample, \cite{Z05} were able to restrict their analyses to volume-limited samples. Our analysis is magnitude-limited, on the other hand, and therefore suffers from projection effects. Our apparent magnitude limit of $r = 21$ can naively be visualized as a large volume containing intrinsically luminous sources and a smaller volume containing intrinsically fainter sources. Given the clustering dependence on luminosity~\citep[see, e.g.,][and references therein]{Z05} the correlation function of these two samples would clearly be different. Since we observe their combination, we naturally see a broader transition region (e.g., compare our Figure~\ref{fig:12} with Figure 22 from \cite{Z05}).

Second, in a 2-point correlation function, only two halo terms need to be used, as two points can either be in the same halo or in two different halos. As $N$ increases, the number of allowable point configurations also increases~\citep[see, e.g.,][]{Ber02}; and these points can be distributed amongst one, two, ..., or $N$ halos. This increase in the number of configurations and in the number of dark matter halos in which they are distributed could explain the increase in the small-angle correlation function slope for late-type galaxies as $N$ increases. We expect to address both of these issues in a future work, by using photometric redshifts to minimize projection effects and quantify luminosity effects on $\bar{\omega}_N$, and by performing an HOD analysis that includes additional halo terms for the higher-order correlation functions.

Figure \ref{fig:ratio} is also very interesting.  Naively, in the absence of higher-order biasing, one would interpret differences in the ratios of $S_3$ and $S_4$ between early- and late-type galaxies as tracing differences in the linear bias of these galaxy types.  Under this paradigm, given that the first-order bias is inversely proportional to the hierarchical amplitudes (see, e.g., Equation 19), we might conclude that we had merely rediscovered the morphology-density relation for galaxies (e.g., \citealt{Dre80}).  As expected under such an interpretation (see, e.g., \citealt{Go03}), we find, in clustered environments, that early-types are more biased than late-types out to some characteristic scale (around $0.5~\Mpch$), on which late-types begin to dominate. However, as we demonstrated in the previous two subsections, our work suggests that it is extremely unlikely that higher-order bias terms are unimportant, particularly when comparing early- and late-type galaxies.

As higher-order bias is important, one might consider an alternative explanation. The blue/red color cut we use should select late-type galaxies as those that are either actively star-forming, or have undergone recent star formation.  It is therefore possible that our classification of a galaxy as late-type is correlated with the likelihood for a galaxy to be in an environment that is conducive to forming stars.  Such unsettled environments should be enhanced by merger events that will be traced by tight groupings of galaxies.  Therefore, on scales less than $0.5~\Mpch$ we find $S_3$ and $S_4$ to be large for late-type galaxies, as their star-forming activity will be traced by compact, unusual configurations of galaxies.  This is particularly true in comparison to early-type galaxies, which, as they are no longer forming stars, are in settled environments that one would not expect to be traced by mergers or tight groupings of galaxies.  

On scales larger than $\sim0.5~\Mpch$, Figure \ref{fig:ratio} suggests a transition towards early-type galaxies having more interesting configurations (i.e., at least 3-4 neighboring galaxies on these scales) than late-types.  This can easily be interpreted, in an ensemble sense, as redder galaxies dominating highly-evolved, cluster-like environments, and bluer galaxies dominating void-like environments (e.g., \citealt{Roj04}).  Thus, our work predicts that in groups of galaxies with characteristic scales $\leqsim~0.5\Mpch$ one should expect merger activity that is triggering star formation, whereas groups with characteristic scales of $\geqsim~0.5\Mpch$ should correlate with more relaxed cluster environments. In future work, we hope to fit Figure \ref{fig:ratio} with models of the relative density profiles that we infer from the local environments of early- and late-type galaxies. 

\section{Conclusions}
We have measured the area-averaged angular correlation functions, $\bar{\omega}_{N}$ ($N \leq 7$), for over 11 million galaxies selected from the SDSS DR3 to have $18 \leq r < 21$, by far the largest sample ever used to perform this type of measurement.  Our large number of objects allowed us to split the full sample, allowing us to perform consistency checks and to inquire into the relationship between galaxy-type and clustering.  Our data samples consisted of four magnitude bins ($18 \leq r < 21$, $18 \leq r < 19$, $19 \leq r < 20$, $20 \leq r < 21$) and a sample of early- and late-type galaxies selected in the magnitude range $18 \leq r < 21$.  

Using our measured $\bar{\omega}_{N}$, we calculated the projected hierarchical amplitudes, $s_{N}$.     Through the use of selection functions empirically determined from the photometric redshifts of galaxies from SDSS DR1, we deprojected our $s_{N}$ and $\bar{\omega}_{2}$ measurements into the real-space amplitudes, $S_{N}$, and the real space volume-averaged 2PCF, $\bar{\xi}_{2}$.  With these values, we were able to test the Gaussianity of the primordial density field and measure galaxy bias as a function of galaxy type.

We checked that the systematic effects of reddening and seeing were not major influences on our measurements.  We found that reddening results in no appreciable effect, even on scales larger than $1^{o}$.  To be safe, we introduced a reddening cut at $A_{g} = 0.3$ ($\sim A_{r} = 0.2$), consistent with the mask suggested by \cite{Scr02}.  Seeing introduces slight structure into the measurements of $s_{N}$, but we found that using a seeing cut at 1\farcs5 minimized the error on our $S_{N}$ measurements.  

Our resulting measurements of $S_{N}$ share consistent amplitudes across each magnitude bin, though there are differences in their structure.  The amplitudes in each magnitude bin are roughly constant at large scales, consistent with the hierarchical model predicted from a Gaussian primordial density field.  Assuming that the first-order bias $b_{1}$ is equal to one when sampling all galaxies, we find significant second and third order bias terms of $c_{2} = -0.24 \pm 0.08$ and $c_{3} = 0.98 \pm 0.89$.  The value of $c_{2}$ is consistent with the findings of \cite{Gaz05}, but slightly inconsistent with those of \cite{H05} and \cite{Pan05}.  Inconsistencies between results should not be surprising based on the differences in the samples and techniques used, especially if separate galaxy data sets sample different galaxy types.  Our marginal result of non-negligible $c_{3}$ is the first that has been reported using SDSS data.  Based on this analysis, we postulate that it is likely that higher-order bias terms are also significant.

There are startling differences between our measurements of $S_{N}$ using early-type galaxies and those using late-types.  At the small scales ($\bar{r} < 1$ $h^{-1}$ Mpc), we find the ratio of $S_{N,early}$ to $S_{N,late}$ increases with scale.  Given that we find higher-order bias terms to be important for early- and late-type galaxies, we attribute this to the fact that late-type galaxies, as we select them, display evidence of recent star formation, and thus might be expected either to be undergoing, or have recently undergone, merger activity. These galaxies exist, therefore, in unsettled environments, traced by interesting overdensities of galaxies on scales of $\leqsim~0.5\Mpch$. Early-type galaxies, on the other hand, typically occupy more settled environments, characteristic of evolved and relaxed galaxy groups and clusters.  Thus, we find that $S_{N,early}$ remains roughly constant as the scale gets smaller, while $S_{N,late}$ shows a dramatic increase.

The ratios $S_{N,early}$ to $S_{N,late}$ are roughly constant at scales $\bar{r} < 1$ $h^{-1}$ Mpc.  We find this can only be explained by large differences in both the linear and higher-order bias.  This result is independent of any assumption on the value of the first-order bias of all galaxies.  If we assume that $b_{1} = 1$ for all galaxies, we find the results summarized in Table 5.  Our result that $c_{2,early} = 0.28 \pm 0.10$ is consistent with \cite{FR05}, suggesting that they were correct in asserting their result was due to the fact that the 2MASS galaxy catalog is dominated by early-type galaxies.

Our results bode well for future measurements using SDSS data.  Using the fifth data release (DR5) of the SDSS, the total area will increase by nearly 70$\%$, allowing for more precise measurements and also for greater precision on large scales, which should help tighten the constraints on the nature of the primordial density field.  Also, DR5 will include photometric redshifts for each galaxy, which will aid in the deprojection process, allow one to study the dependence of clustering on luminosity, and allow one to explore any cosmic evolution in the hierarchical amplitudes.  We anticipate that this information will aid in producing a clearer picture of the nature of galaxy bias, both for all galaxies and as a function of galaxy type.

\acknowledgements

AJR, RJB and ADM acknowledge support from NASA through grants NAG5-12578 and NAG5-12580, Microsoft Research, and the NSF PACI Project.  The authors made extensive use of the storage and computing facilities at the National Center for Supercomputing Applications and thank the technical staff for their assistance in enabling this work.

We thank Ani Thakar and Jan Van den Berg for help with obtaining a copy of the SDSS DR3 databases and Ryan Scranton for his extensive help with implementing \SDSSpixb.  We also would like to thank Scott Dodelson, Joshua Frieman, Istvan Szapudi, and an anonymous referee for comments and suggestions that improved the paper.

Funding for the creation and distribution of the SDSS Archive has been provided by the Alfred P. Sloan Foundation, the Participating Institutions, the National Aeronautics and Space Administration, the National Science Foundation, the U.S. Department of Energy, the Japanese Monbukagakusho, and the Max Planck Society. The SDSS Web site is http://www.sdss.org/.

The SDSS is managed by the Astrophysical Research Consortium (ARC) for the Participating Institutions. The Participating Institutions are The University of Chicago, Fermilab, the Institute for Advanced Study, the Japan Participation Group, The Johns Hopkins University, the Korean Scientist Group, Los Alamos National Laboratory, the Max-Planck-Institute for Astronomy (MPIA), the Max-Planck-Institute for Astrophysics (MPA), New Mexico State University, University of Pittsburgh, University of Portsmouth, Princeton University, the United States Naval Observatory, and the University of Washington.

\appendix
\section{From Connected Moments to N-point correlations}
\label{app:a}

The N-point correlations, $\omega_{N}$, can be derived from the connected moments, $m_{N}$
via the connected central moment, $\mu_{N}$, as described in \citet{Gaz94}.  Here, we present the relevant formulae for $N \leq 7$:

\begin{equation}
\omega_{N} = \frac{k_{N}}{\bar{n}^{N}}
\end{equation}
and
\begin{equation}
\mu_{2} = m_{2};~~
 k_{2} = \mu_{2} - \bar{n}; 
\end{equation}

\begin{equation}
\mu_{3} = m_{3};~~
 k_{3} = \mu_{3} -3k_{2}- \bar{n}; 
\end{equation}

\begin{equation}
\mu_{4} = m_{4} -3m_{2}^{2};~~
 k_{4} = \mu_{4} -7k_{2}-6k_{3}- \bar{n}; 
\end{equation}

\begin{equation}
\mu_{5} = m_{5} -10m_{3}m_{2};~~
 k_{5} = \mu_{5} -15k_{2}-25k_{3}- 10k_{4}-\bar{n}; 
\end{equation}

\begin{equation}
\mu_{6} = m_{6} -15m_{4}m_{2} - 10m_{3}^{2} + 30m_{2}^{3};~~
k_{6} = \mu_{6} -31k_{2}-90k_{3}- 64k_{4}-15k_{5}-\bar{n}; 
\end{equation}

\begin{equation}
\mu_{7} = m_{7} -21m_{5}m_{2} - 35m_{4}m_{3} + 210m_{3}m_{2}^{2};~~
k_{7} = \mu_{7}- 63k_{2}-301k_{3}- 350k_{4}-140k_{5}-21k_{6}-\bar{n}; 
\end{equation}
\tablewidth{0pt}

\clearpage
\begin{deluxetable}{lccccc}
%\multicolumn{1}{c}{ 1 column } $N$   \multicolumn{2}{c}{ 2 columns }  $\alpha_{N}$ 
\tablecaption{Measured values for the exponent, $\alpha_{N}$, and the amplitude at 1 $h^{-1}$ Mpc, $B_{N}$, assuming power-law behavior for hierarchical amplitudes $S_{N}$, $N \leq 7$, in the bin $18 \leq r < 21$.} 
\tablecolumns{6}
\tablehead{ \colhead{} & \multicolumn{2}{c}{$0.2 < \bar{r} < 2.0$ $h^{-1}$ Mpc} &  \colhead{} & \multicolumn{2}{c}{$2.0 < \bar{r} < 10$ $h^{-1}$ Mpc} \\ \cline{2-3} \cline{5-6} \\
\colhead{$N$} & \colhead{$\alpha_{N}$} & \colhead{$B_{N}$} &  \colhead{} & \colhead{$\alpha_{N}$} & \colhead{$B_{N}$}
}
\startdata
3 & $0.065 \pm 0.009$ & $4.66 \pm 0.06$ & ~ & $-0.31 \pm 0.04$ & $5.15_{-0.24}^{+0.19}$\\
4 & $0.037 \pm 0.036$ & $41.2 \pm 2.1$ & ~ &  $-0.73 \pm 0.09$ & $67.4 \pm 7.0$\\
5 & $0.36 \pm 0.03$ & $810 \pm 40$ & ~ &   $-0.69_{-0.21}^{+0.17}$ & $990_{-210}^{+100}$\\
6 & $0.28_{-0.20}^{+0.26}$ & $(1.36_{-0.34}^{+0.37}) \times 10^4$ &~ &   $-0.27_{-0.38}^{+0.24}$ & $(8.8_{-3.5}^{+3.6}) \times 10^3$\\
7 & $0.40_{-0.35}^{+0.57}$ & $(3.3 \pm 1.7) \times 10^5$ & ~ & $-0.87_{-0.98}^{+0.40}$ & $(3.1\pm 2.3) \times 10^5$
\enddata
  \label{T:slope1}
\end{deluxetable}

\begin{deluxetable}{lccccc}
\tablecaption{Measured values for the exponent, $\alpha_{N}$, and the amplitude at 1 $h^{-1}$ Mpc, $B_{N}$, assuming power-law behavior for hierarchical amplitudes $S_{N}$, $N \leq 7$, in the bin $18 \leq r < 19$.  }
\tablecolumns{6}
\tablehead{ \colhead{} & \multicolumn{2}{c}{$0.2 < \bar{r} < 2.0$ $h^{-1}$ Mpc} &  \colhead{} & \multicolumn{2}{c}{$2.0 < \bar{r} < 10$ $h^{-1}$ Mpc} \\ \cline{2-3} \cline{5-6} \\
\colhead{$N$} & \colhead{$\alpha_{N}$} & \colhead{$B_{N}$} &  \colhead{} & \colhead{$\alpha_{N}$} & \colhead{$B_{N}$}
}
\startdata
3 & $0.03 \pm 0.03$ & $3.86 \pm 0.03$ & ~ & $-0.13 \pm 0.05$ & $3.50 \pm 0.20$\\
4 & $0.07 \pm 0.10$ & $34.8 \pm 2.7$ & ~ &  $-0.22 \pm 0.15$ & $26.7 \pm 3.8$\\
5 & $0.18_{-0.27}^{+0.36}$ & $417 \pm 94$ & ~ &   $-0.12 _{-0.23}^{+0.18}$ & $258 \pm 55$\\
6 & $0.28_{-0.32}^{+1.36}$ & $(5.32 \pm 2.43) \times 10^3$ &~ &   $0.00_{-0.50}^{+0.31}$ & $(2.68 \pm 1.05) \times 10^3$\\
7 & $0.16_{-0.95}^{+13.88}$ & $(6.5_{-6.4}^{+5.5}) \times 10^5$ & ~ & $0.1_{-1.2}^{+0.5}$ & $(2.9 \pm 2.0) \times 10^4$
\enddata

 \label{T:slope2}
\end{deluxetable}

\begin{deluxetable}{lccccc}

 \tablecaption{Average $s_{N}$ and $S_{N}$ taken for data between 0.2 and 10 $h^{-1}$ Mpc for galaxies with $18<r<19$ and $18 < r <21$}
\tablehead{ 
\colhead{} & \multicolumn{2}{c}{$18 < r < 21$} &  \colhead{} & \multicolumn{2}{c}{$18 < r < 19$ } \\
 \cline{2-3} \cline{5-6}\\

\colhead{$N$} & \colhead{$s_{N}$} & \colhead{$S_{N}$} &  \colhead{} & \colhead{$s_{N}$} & \colhead{$S_{N}$}
}\startdata
3 & 5.38 $\pm$ 0.03 & 4.72 $\pm$ 0.03 &~&  5.69 $\pm$ 0.08 & 4.24 $\pm$ 0.06\\
4 & 67.4 $\pm$ 2.1 & 49.9 $\pm$ 1.5 &~& 66.4 $\pm$ 4.8 & 34.8 $\pm$ 2.5\\
5 & 1040 $\pm$ 70 & 635 $\pm$ 42 &~& 1400 $\pm$ 90 & 492 $\pm$ 32\\
6 & (1.71 $\pm$ 0.21)x10$^{4}$ & (8.51 $\pm$ 1.05)x10$^{3}$ &~& (3.87 $\pm$ 0.16)x10$^{4}$ & (8.90 $\pm$ 0.37)x10$^{3}$\\
7 & (4.44 $\pm$ 0.80)x10$^{5}$ & (1.78 $\pm$ 0.32)x10$^{5}$ &~& (8.38 $\pm$ 0.20)x10$^{5}$ & (1.23$\pm$ 0.29)x10$^{5}$\\

\enddata
 \label{T:ave}
\end{deluxetable}

 \begin{deluxetable}{lccccc}

\centering
\tablecaption{Average $s_{N}$ and $S_{N}$ taken for data between 1 and 10 $h^{-1}$ Mpc for early- and late-type galaxies with $18<r<21$ } 
\tablehead{ 
\colhead{} & \multicolumn{2}{c}{Early-type} &  \colhead{} & \multicolumn{2}{c}{Late-type } \\
 \cline{2-3} \cline{5-6}\\
\colhead{$N$} & \colhead{$s_{N}$} & \colhead{$S_{N}$} &  \colhead{} & \colhead{$s_{N}$} & \colhead{$S_{N}$}

 }
 \startdata
3 &  4.76 $\pm$ 0.11 & 4.24 $\pm$ 0.10 &~& 2.83 $\pm$ 0.19 & 2.32 $\pm$ 0.16\\
4 & 47.6 $\pm$ 3.3 & 36.6 $\pm$ 2.5 &~& 22.5 $\pm$ 1.9 &  14.4 $\pm$ 1.2\\
5 & 1780 $\pm$ 40 & 1160 $\pm$ 30 &~& 149 $\pm$ 78 & 72.7 $\pm$ 39\\
6 & (4.96 $\pm$ 0.13)x10$^{4}$ & (2.68 $\pm$ 0.07)x10$^{4}$ &~& (4.73 $\pm$ 1.47)x10$^{3}$ & (1.73 $\pm$ 0.54)x10$^{3}$\\
7 & (1.75 $\pm$ 0.64)x10$^{6}$ & (7.78 $\pm$ 2.84)x10$^{5}$ &~& (1.1 $\pm$ 4.6)x10$^{5}$ & (0.3 $\pm$ 1.3)x10$^{4}$\\

\enddata
  \label{T:ave2 }
\end{deluxetable}

\begin{deluxetable}{lccc}

\centering
\tablecaption{The determined values of first-, second-, and third-order bias for late-type, early-type, and all galaxies, assuming that $b_{1}$ = 1 for all galaxies.} 
\tablehead{
\colhead{Sample} &
\colhead{$b_{1}$} &
\colhead{$c_{2}$} &
\colhead{$c_{3}$}
}
\startdata
All, $18<r<21$ &  1 & -0.24 $\pm$ 0.08 & 0.98 $\pm$ 0.89 \\
All, $18<r<19$ &  1 & -0.31 $\pm$ 0.08 & 1.23 $\pm$ 0.94 \\
Early-type, $18<r<21$ & 1.36 $\pm$ 0.04 &  0.30 $\pm$ 0.10 & 1.46 $\pm$ 1.37\\
Late-type, $18<r<21$ & 0.81 $\pm$ 0.03 &  -0.71 $\pm$ 0.07 & 0.63 $\pm$ 1.04

\enddata
\label{T:bias}
\end{deluxetable}

\clearpage
\begin{figure}
\plotone{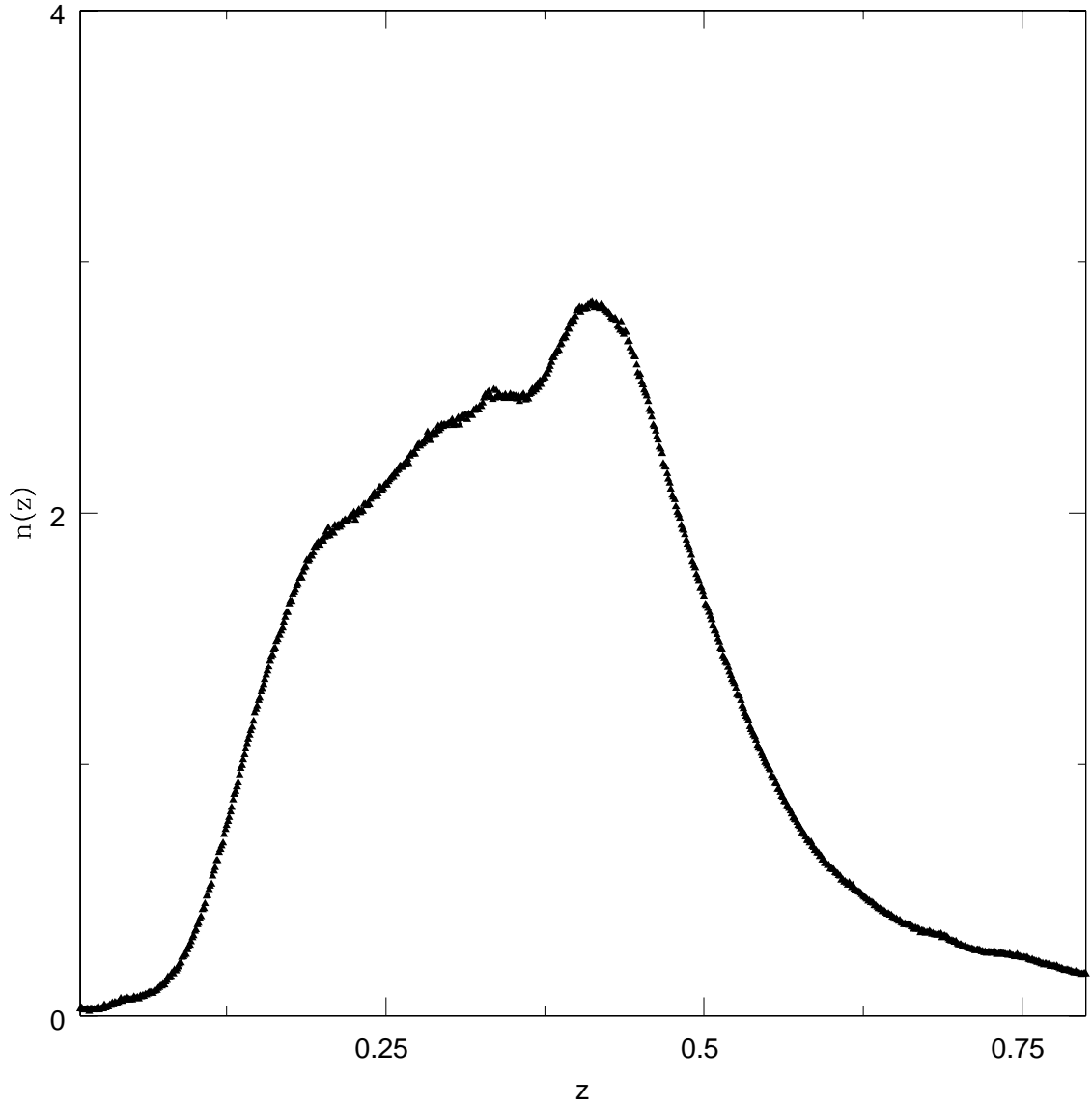}
\caption{Normalized number of galaxies, $n(z)$, in redshift bins $\Delta z = 10^{-3}$ determined from photometric redshifts of galaxies with $18 \leq r < 21$.  This curve defines the redshift selection function we use in our deprojections from angular to real-space for hierarchical amplitudes the magnitude bin $18 \leq r < 21$.  Similar curves are constructed for the other magnitude bins.}
\label{fig:1}
\end{figure}
\clearpage
\begin{figure}
\plotone{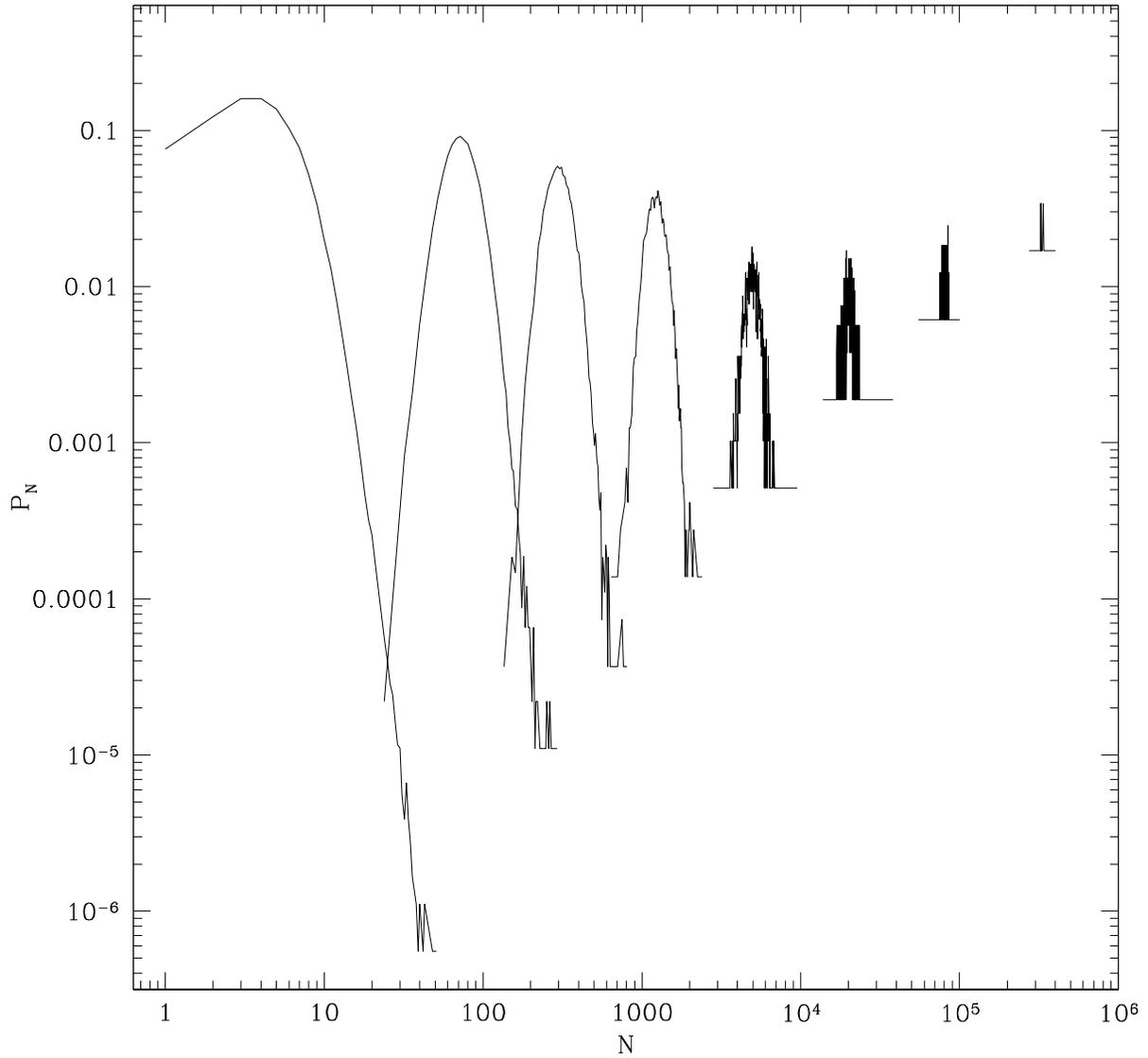}
\caption{The probability of having $N$ counts in a cell for resolutions corresponding to the angular scales 0.02$^{0}$,0.08$^{0}$,0.16$^{0}$,0.3$^{0}$,0.6$^{0}$,1.3$^{0}$,2.6$^{0}$, and 5.2$^{0}$ (from left to right). The discretization in these large angle curves indicates make precise measurements at scales larger than a about a degree.}
\label{fig:2}
\end{figure}
\clearpage
\begin{figure}
\plotone{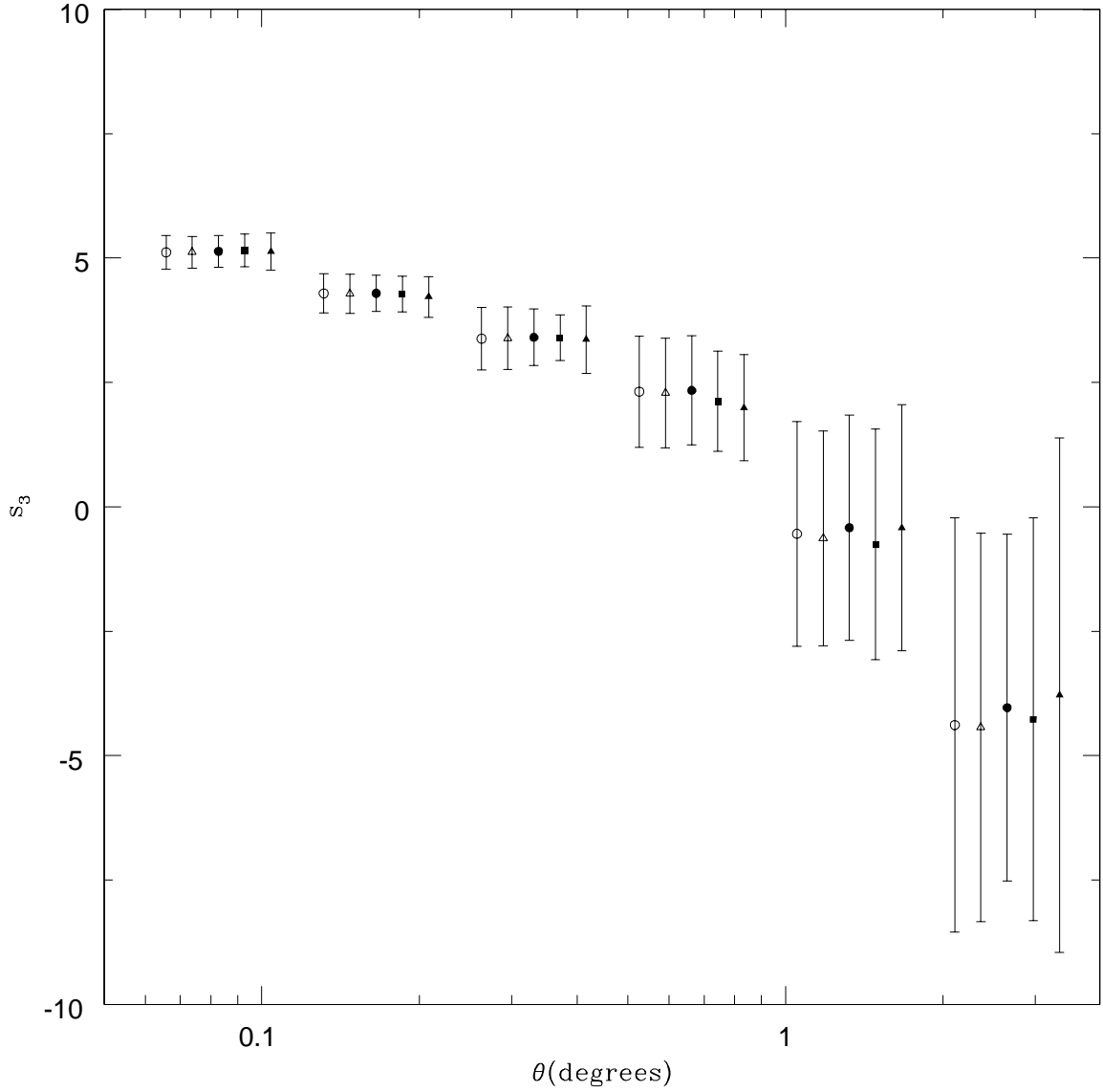}
\caption{The $s_{3}$ hierarchical amplitude for 18 $\leq r<$ 21 galaxies with $g$-band reddening cuts at 0.35 (open circles); 0.3 (open squares); 0.25 (filled circles); 0.2 (filled squares); 0.15 (filled triangles).  The $\theta$ values have been shifted for each seeing cut for clarity.  We see little change in the amplitudes as a function of the reddening cut.}
\label{fig:3}
\end{figure}
\clearpage

\begin{figure}
\plotone{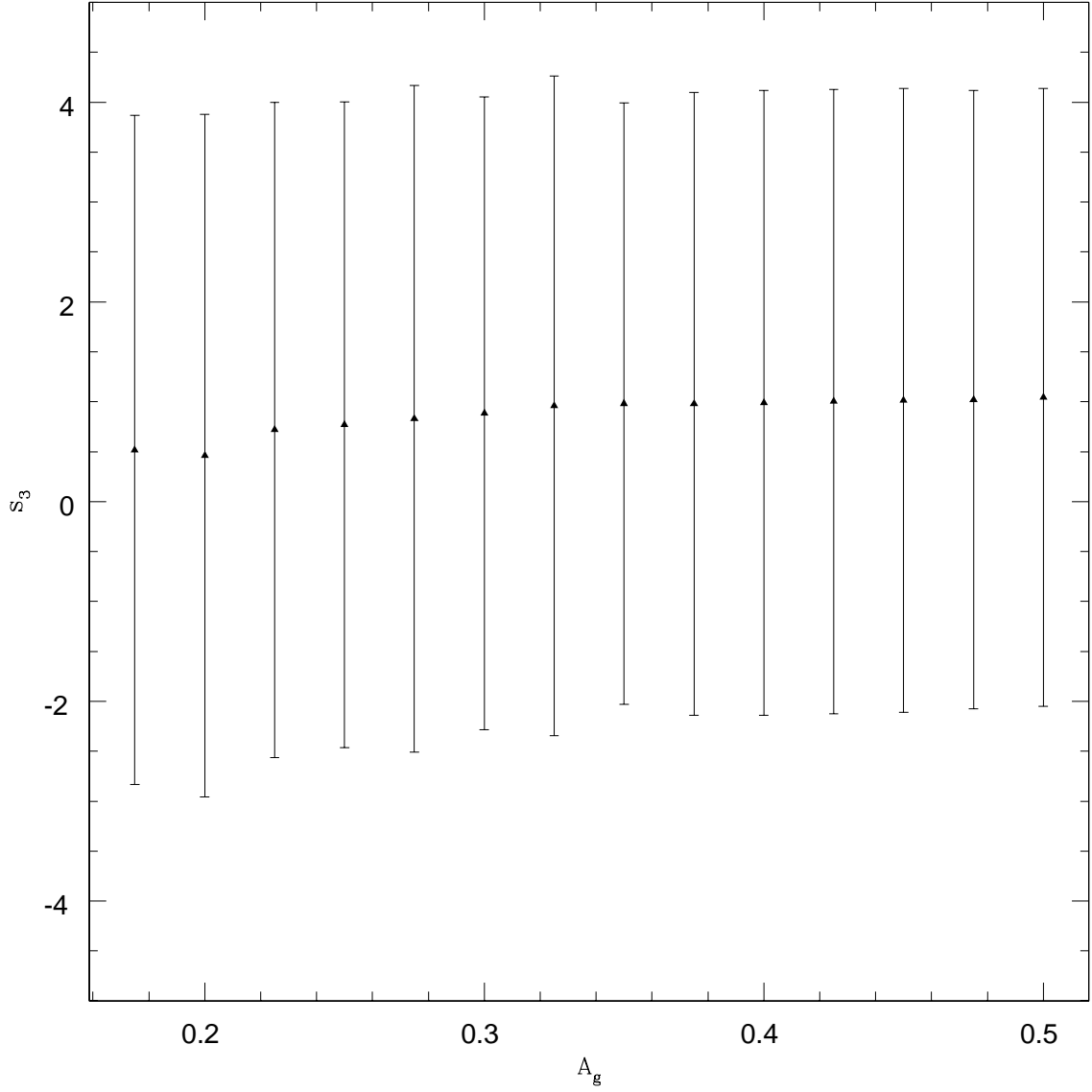}
\caption{The $s_{3}$ hierarchical amplitude at 1.3$^0$ versus the $g$-band reddening cut that is applied.  We determine the preferred cut to be at $A_{g} = 0.3$.}
\label{fig:4}
\end{figure}
\clearpage
\begin{figure}
\plotone{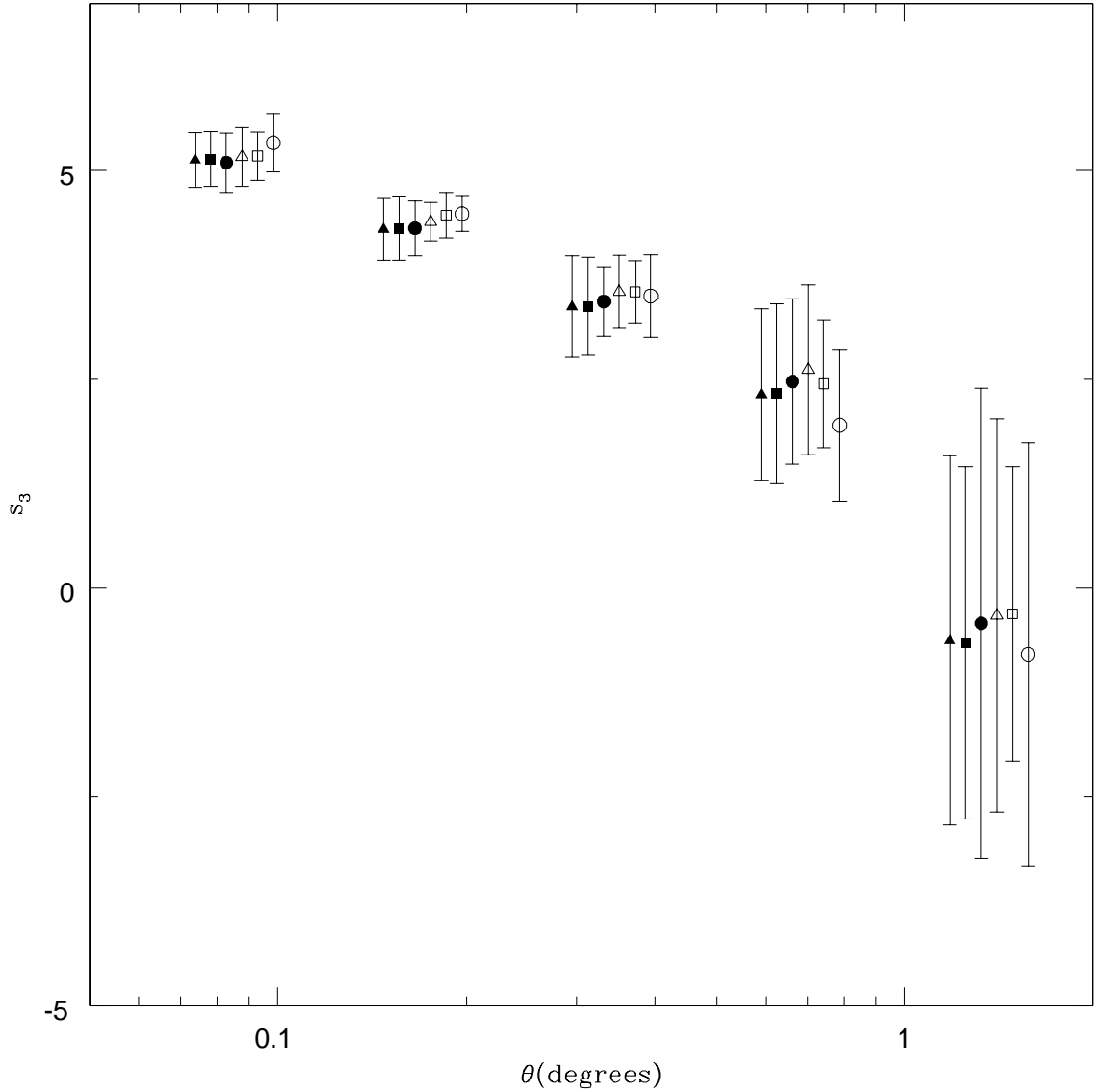}
\caption{The $s_{3}$ hierarchical amplitudes of $18 \leq r < 21$ galaxies with seeing cuts at 1\farcs4 (open circles); 1\farcs5 (open squares); 1\farcs6 (open triangles); 1\farcs7 (filledcircles); 1\farcs9 (filled squares); and 2\farcs2 (filled triangles).  The $\theta$ values have been shifted for each seeing cut for clarity.  Seeing does introduce minor structure into our measurements, at a level below our errors.  The error on the measurement is minimized at our preferred value of 1\farcs5 for most scales.}
\label{fig:5}
\end{figure}
\clearpage
\begin{figure}
\plotone{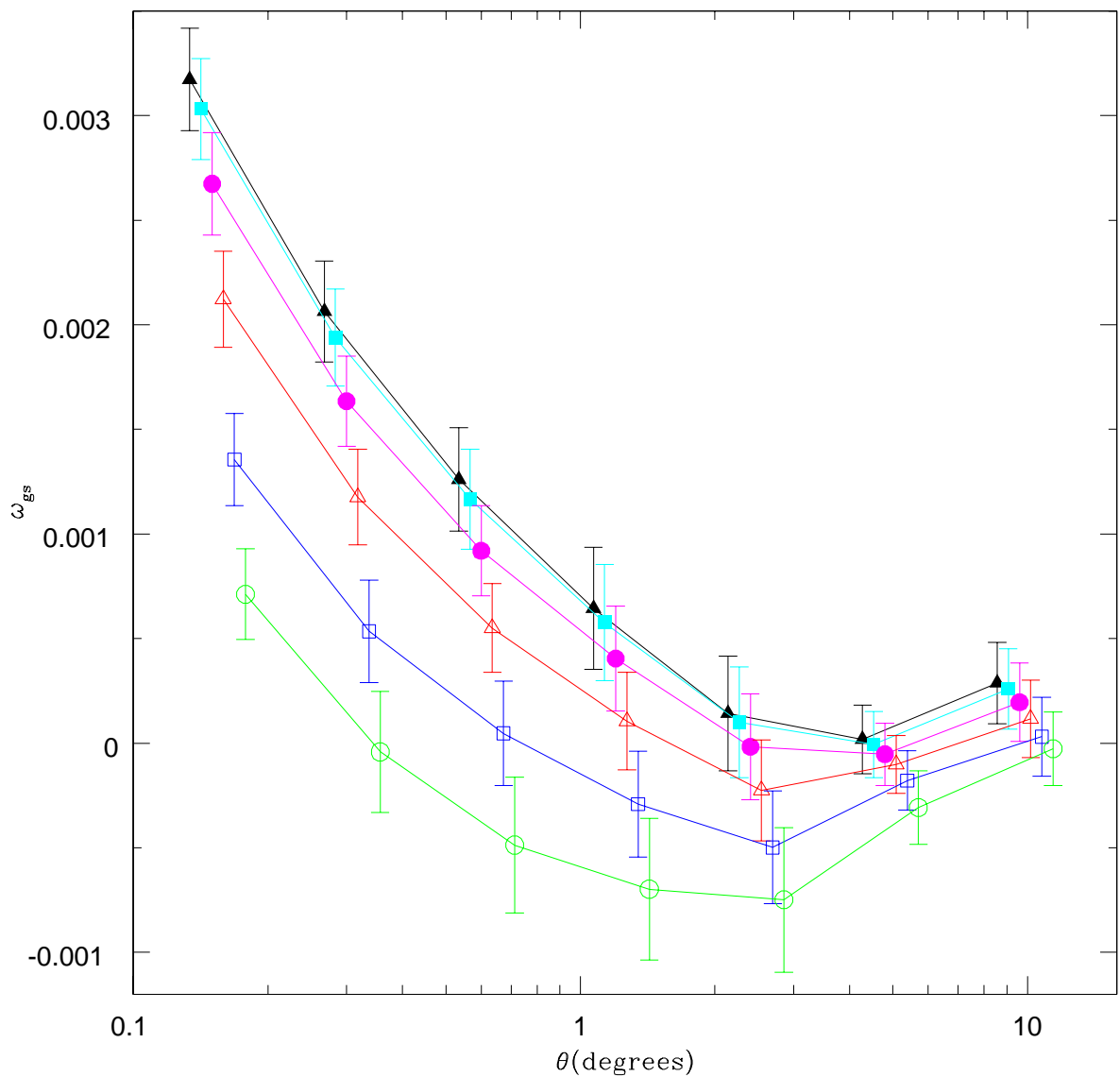}
\caption{Galaxy-seeing cross-correlations with seeing cuts at 1\farcs4 (green, open circles); 1\farcs5 (blue, open squares); 1\farcs6 (red, open triangles); 1\farcs7 (magenta, filled circles); 1\farcs8 (cyan, filled squares); and 1\farcs9 (black, filled triangles) for $18 \leq r < 21$ galaxies.  The $\theta$ values have been shifted slightly for each seeing cut for clarity.  The cross-correlation for the cut at out preferred seeing cut of 1\farcs5 is the most consistent with zero.}
\label{fig:6}
\end{figure}
\clearpage
\begin{figure}
\plotone{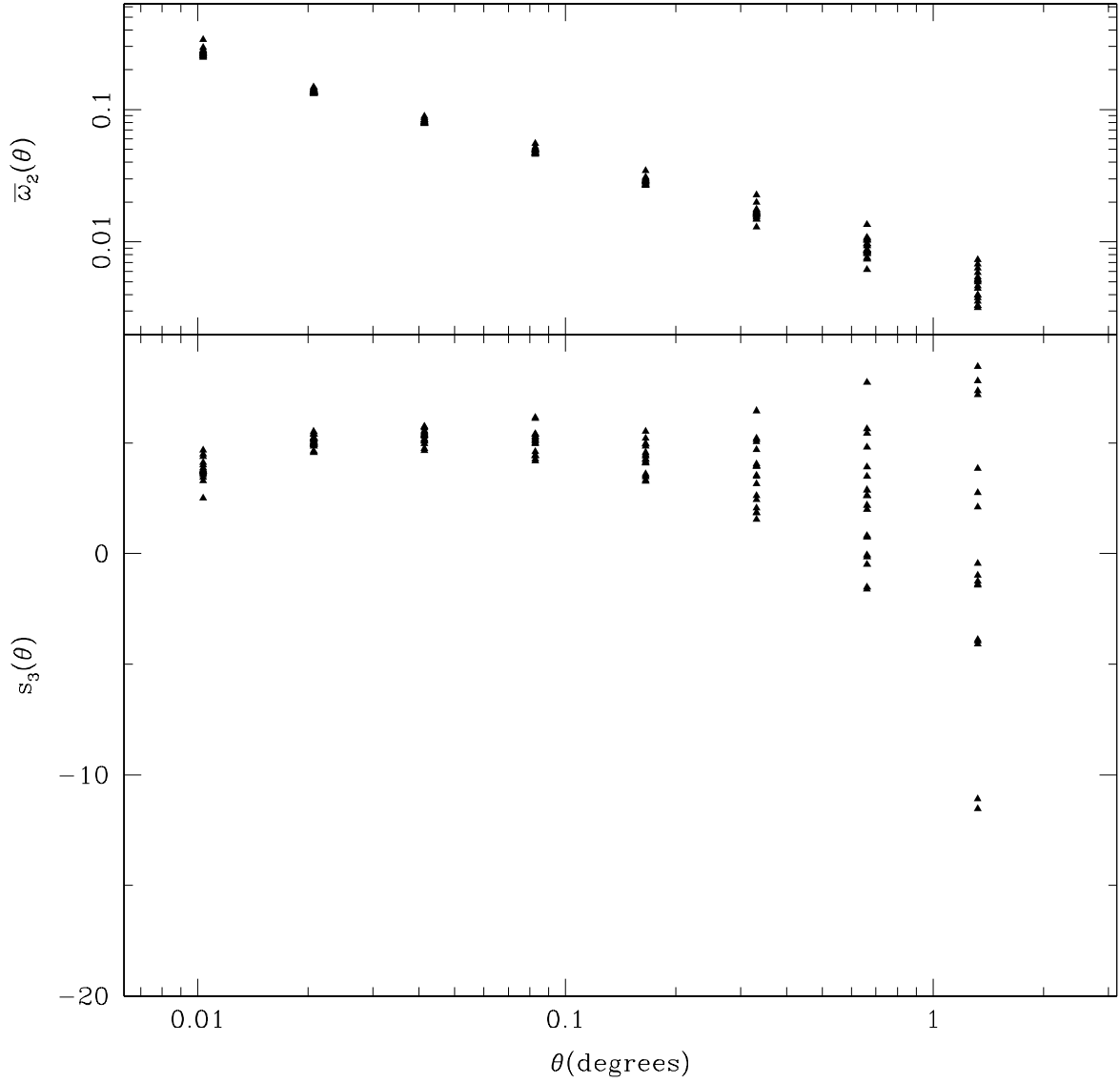}
\caption{The values $\bar{\omega}_{3}(\theta)$ and $s_{3}(\theta)$ for each SDSS data stripe.  There is minimal variation between stripes and no one stripe dominates the measurements.}
\label{fig:7}
\end{figure}
\clearpage
\begin{figure}
\plotone{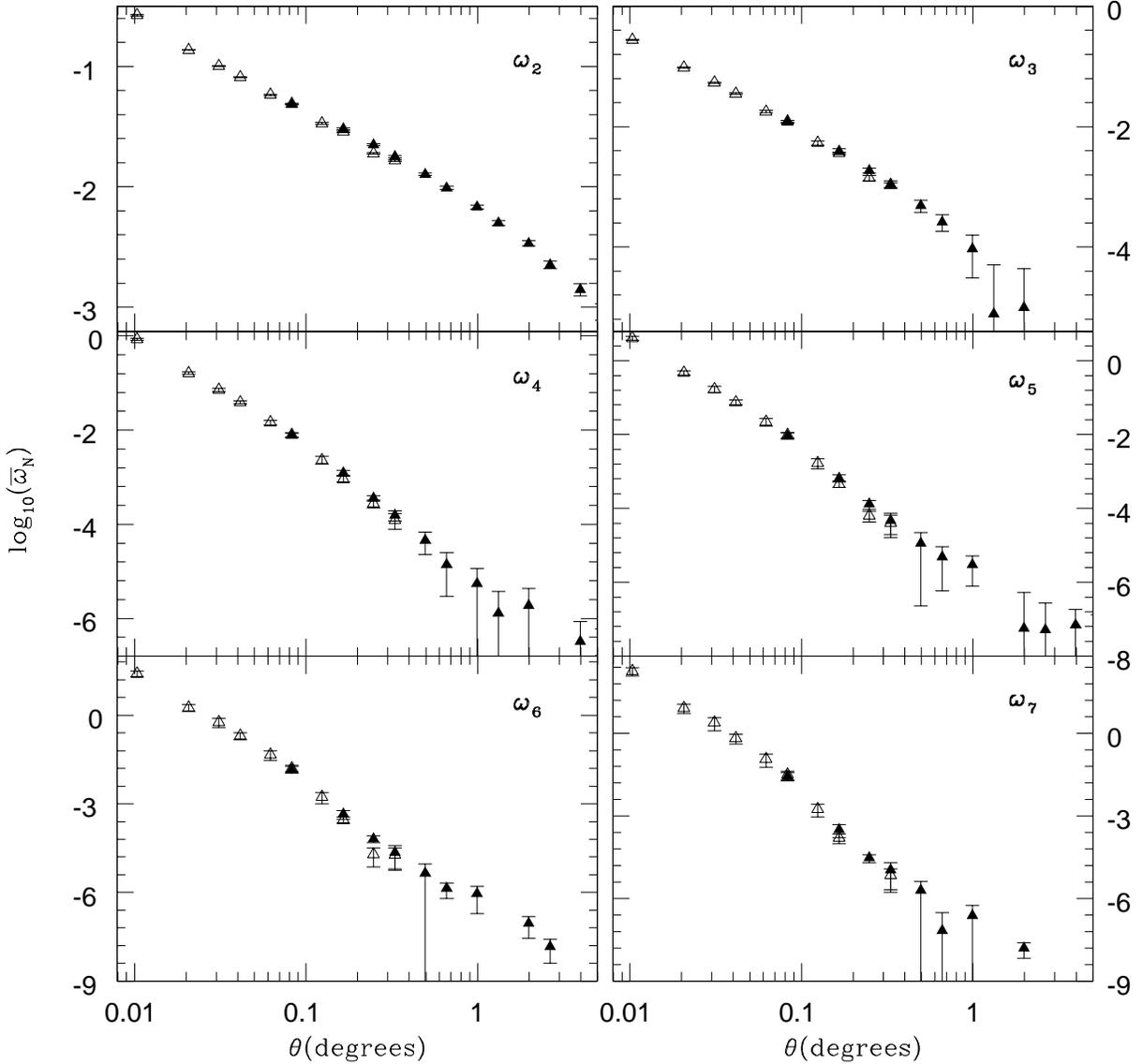}
\caption{The area-averaged angular N-point correlation functions for $18 \leq r < 21$  galaxies for $N$ = 2,.....,7 (left to right, top to bottom).  Note the logarithmic scaling.  The power-law behavior is observed for all $N$, consistent with the hierarchical model.  The open triangles designate data calculated using the striped method, while solid triangles represent the data calculated using the \SDSSpixt method.  There is no measurement using the \SDSSpixt method at 0.12 degrees because this scale is not an integer multiple of its base resolution scale (0.08 degrees).}
\label{fig:8}
\end{figure}

\clearpage

\begin{figure}[hbtp]
\plotone{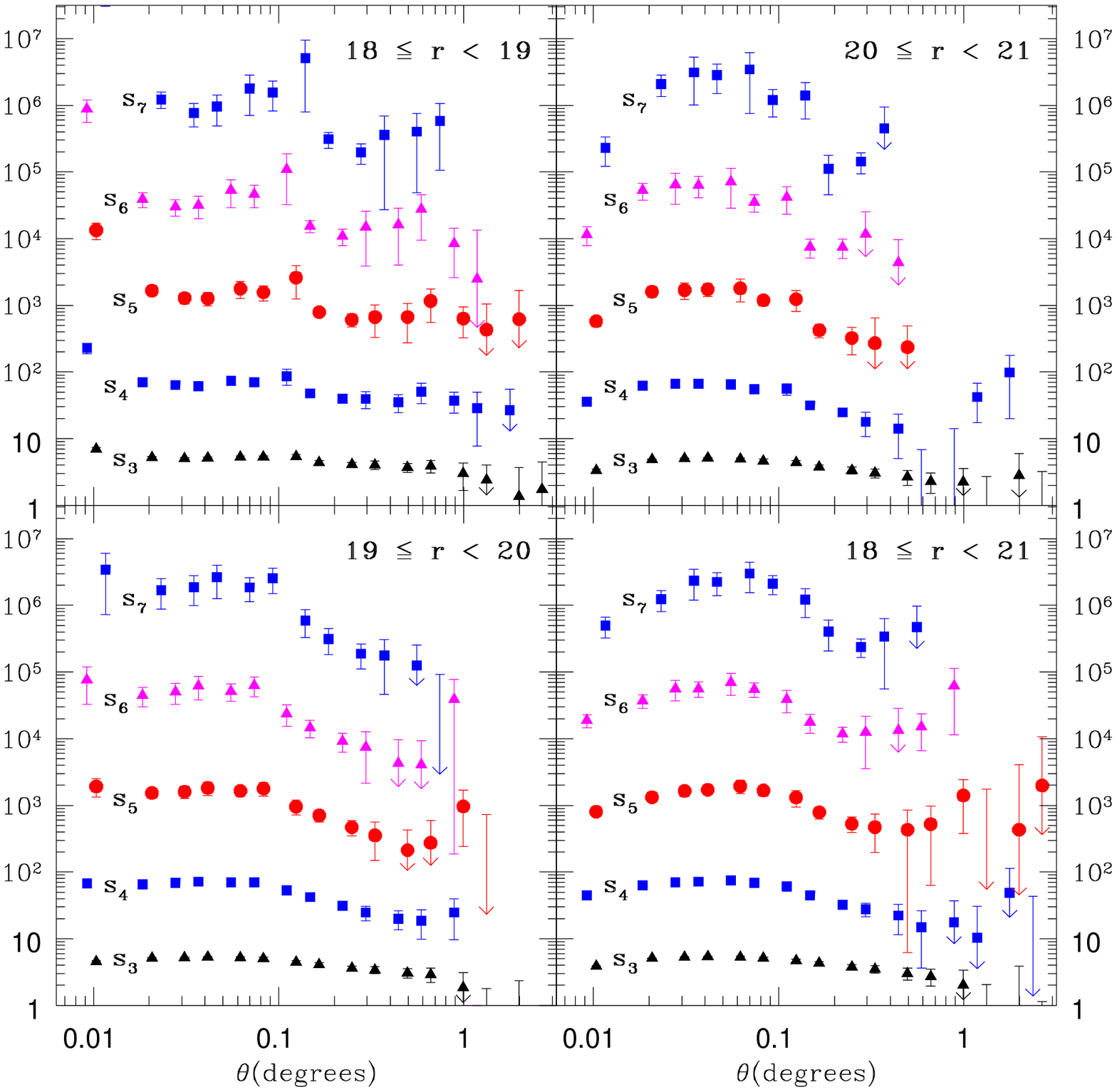}
\caption{The hierarchical amplitudes $s_{3}$ through $s_{7}$ for galaxies with $18 \leq r < 19$ (top left), $19 \leq r < 20$ (bottom left), $20 \leq r < 21$ (top right), and $18 \leq r < 21$ (bottom right). The $\theta$ values for each amplitude have been shifted and data with extremely large errors are omitted for clarity.  The amplitudes are roughly constant, consistent with the hierarchical model.}
\label{fig:9}
\end{figure}

\clearpage

\begin{figure}[hbtp]
\plotone{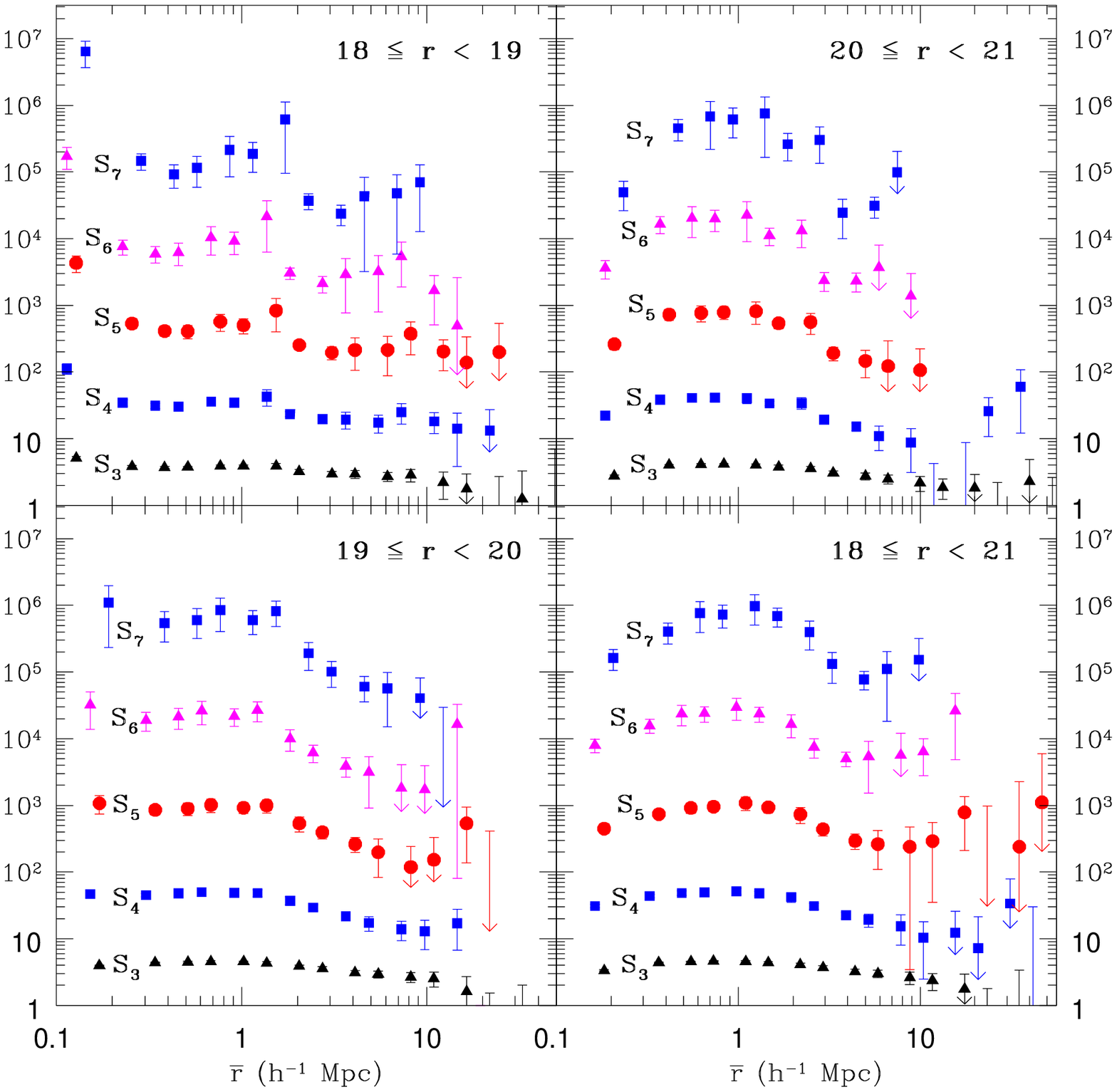}
\caption{The hierarchical amplitudes $S_{3}$ through $S_{7}$ for galaxies with $18 \leq r < 19$ (top left), $19 \leq r < 20$ (bottom left), $20 \leq r < 21$ (top right), and $18 \leq r < 21$ (bottom right). The $\bar{r}$ values for each amplitude have been shifted and data with extremely large errors are omitted for clarity.  The amplitudes are roughly constant, consistent with the hierarchical model.}
\label{fig:10}
\end{figure}

\clearpage

\begin{figure}[hbtp]
\plotone{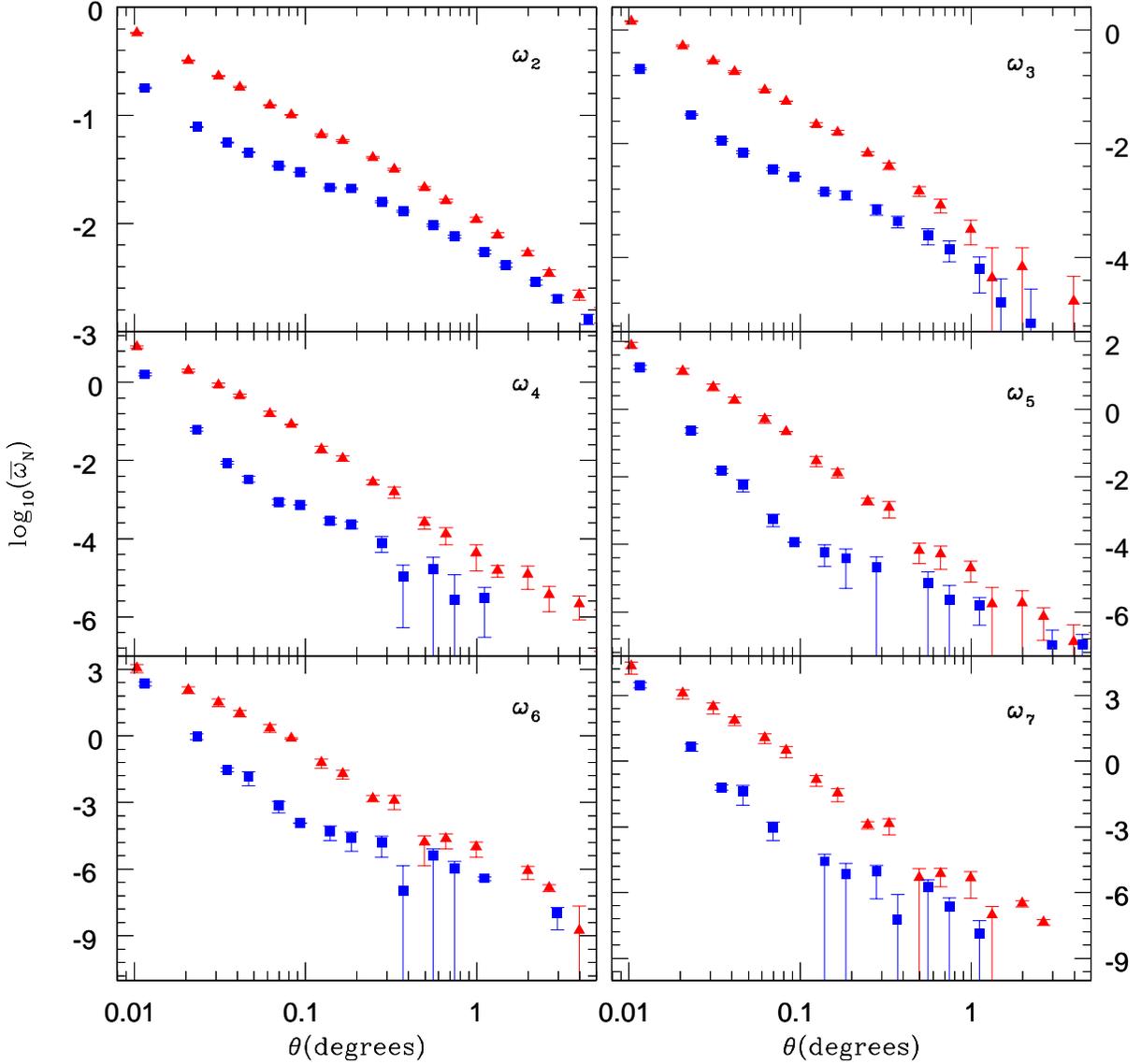}
\caption{The area-averaged angular N-point correlation functions for $18 \leq r < 21$ early-type (red, triangles) and late-type (blue, squares) galaxies, for $N$ = 2,.....,7 (left to right, top to bottom).  Note the logarithmic scaling, and that the $\theta$ values of the early-type data have been shifted to slightly lower values for clarity.  Clear differences between the two galaxy types are evident in the slopes and amplitudes.}
\label{fig:12}
\end{figure}

\clearpage

\begin{figure}[hbtp]
\plotone{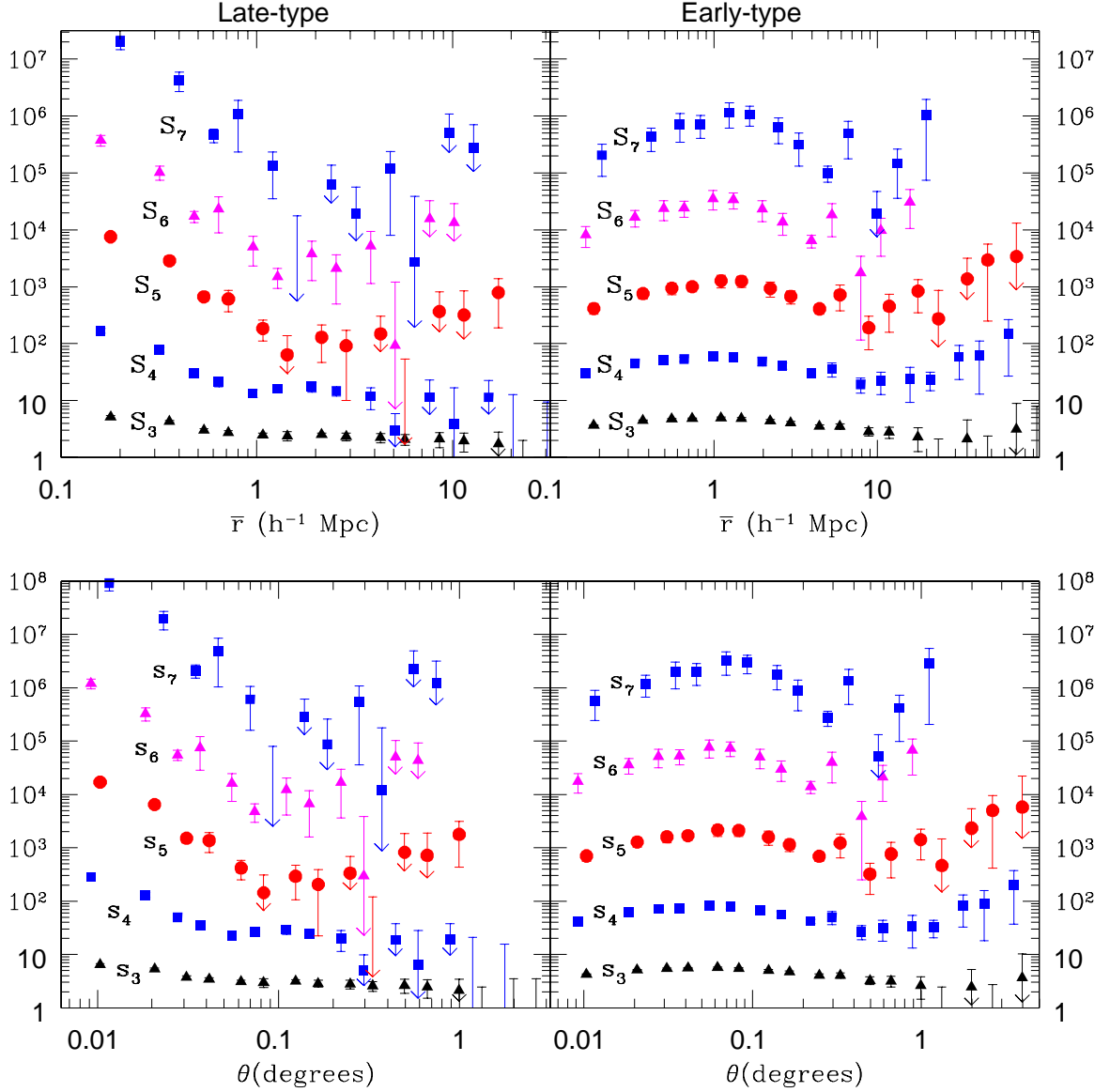}
\caption{The hierarchical amplitudes $s_{3}$ through $s_{7}$ (bottom) and $S_{3}$ through $S_{7}$ (top) for late-type (left) and early-type (right) galaxies. The $\theta$ values for $s_{4}$ and $s_{7}$ have been shifted and data with extremely large errors are omitted for clarity.  At large scales, the early-types have significantly larger amplitudes than the late-types, implying a significant difference in bias between the two samples.}
\label{fig:13}
\end{figure}

\clearpage

\begin{figure}[hbtp]
\plotone{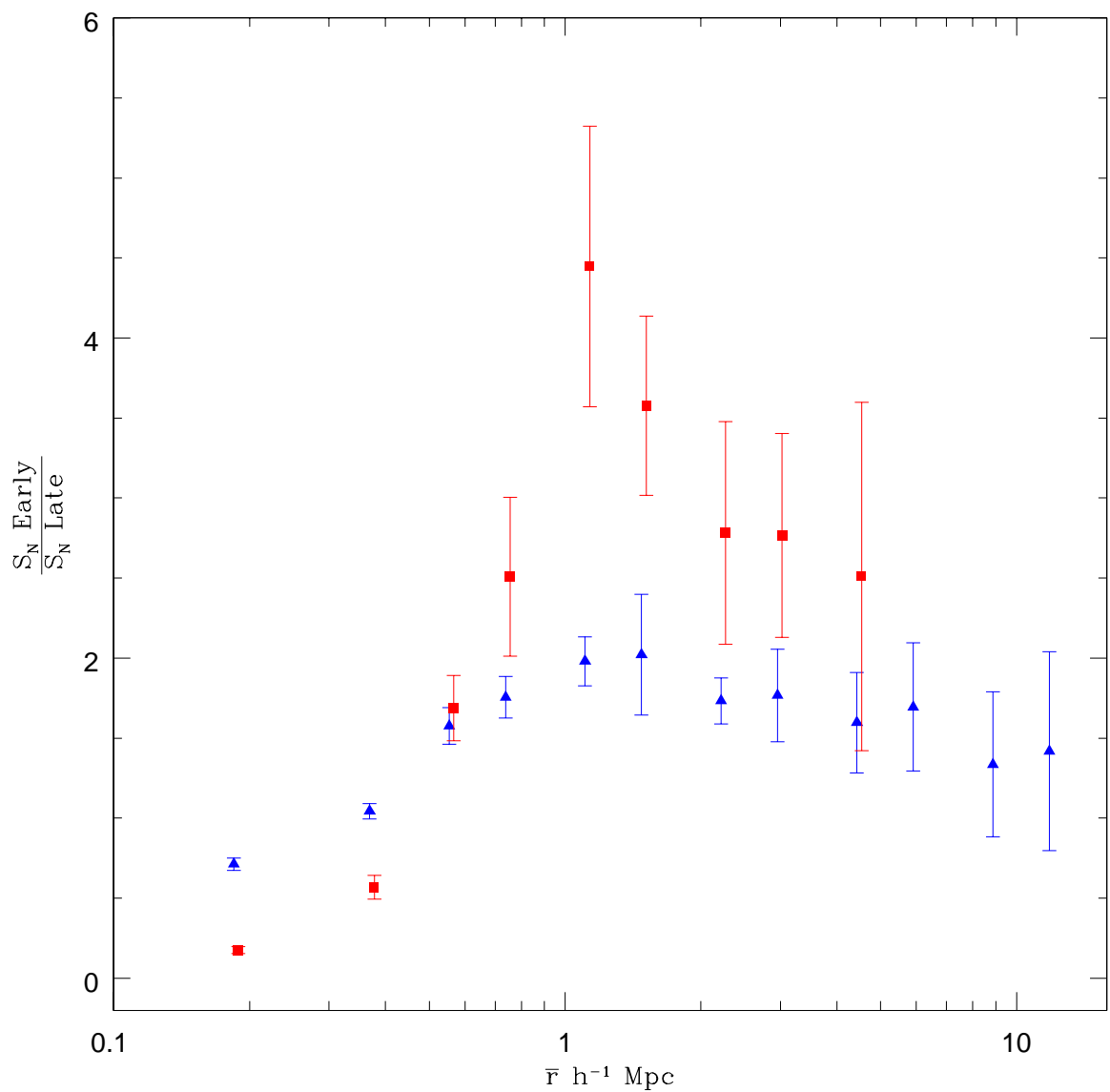}
\caption{The ratios of $S_{3,early}$ to $S_{3,late}$ (blue, triangles) and $S_{4,early}$ to $S_{4,late}$ (red, squares) for galaxies in the magnitude range $18 \leq r < 21$.  Data with large errors are omitted for clarity.  These at large scales, the ratios indicate that there is a significant difference in bias between early- and late-type galaxies.  The ratios at small scales may be indicative of recent merger activities  in late-type galaxies.}
\label{fig:ratio}
\end{figure}

\clearpage

\end{document}